\documentclass[onecolumn,authoryear]{els-mrw} 

\usepackage{amsmath,amssymb,amsfonts,amsthm,makeidx,graphicx}
\usepackage{txfonts}
\usepackage{helvet}

\usepackage[dvipsnames]{xcolor} 
\usepackage{mathtools} 
\usepackage[colorlinks]{hyperref} 
\usepackage{textgreek} 
\usepackage{upgreek} 

\long\def\symbolfootnote[#1]#2{\begingroup%
\def\thefootnote{\fnsymbol{footnote}}\footnote[#1]{#2}\endgroup} 
\def\aj{AJ}
\def\araa{ARA\&A}
\def\aap{A\&A}
\def\icarus{Icarus}
\def\nat{Nature}
\def\ssr{Space~Sci.~Rev.}

\newcommand{\presecContentsEntry}[1]{\noindent\nameref{#1} \hfill \pageref{#1} \newline}
\newcommand{\secContentsEntry}[1]{\noindent\ref{#1}. \nameref{#1} \hfill \pageref{#1} \newline}
\newcommand{\subsecContentsEntry}[1]{\forceindent\ref{#1}. \nameref{#1} \hfill \pageref{#1}\newline}
\newcommand{\forceindent}{\leavevmode{\parindent=2em\indent}}

\newcommand{\horizontalLine}[1]{\noindent\textcolor{darkgray}{\rule{\textwidth}{1pt} }}

\mathchardef\shorthyphen="2D                                            
\newcommand{\mEarth}{\; {\rm M_\oplus}}                         

\begin{document}

\chapter{The Solar System: structural overview, origins and evolution}\label{chap1}

\author[1]{Sean N. Raymond}%

\address[1]{\orgname{Laboratoire d'Astrophysique de Bordeaux}, \orgdiv{CNRS, Universit{\'e} de Bordeaux}, \orgaddress{Batiment B18N, All{\'e}e Geoffroy Saint Hilaire, 33615 Pessac, France}}


\maketitle


\begin{abstract}[Abstract]
Understanding the origin and long-term evolution of the Solar System is a fundamental goal of planetary science and astrophysics.  This chapter describes our current understanding of the key processes that shaped our planetary system, informed by empirical data such as meteorite measurements, observations of planet-forming disks around other stars, and exoplanets, and nourished by theoretical modeling and laboratory experiments.  The processes at play range in size from microns to gas giants, and mostly took place within the gaseous planet-forming disk through the growth of mountain-sized planetesimals and Moon- to Mars-sized planetary embryos. A fundamental shift in our understanding came when it was realized (thanks to advances in exoplanet science) that the giant planets' orbits likely underwent large radial shifts during their early evolution, through gas- or planetesimal-driven migration and dynamical instability. The characteristics of the rocky planets (including Earth) were forged during this early dynamic phase. Our Solar System is currently middle-aged, and we can use astrophysical tools to forecast its demise in the distant future.
\end{abstract}

\vspace{5mm}

\begin{keywords}
   Solar System formation, Solar System evolution, planet formation, circumstellar disks, meteorites, exoplanets, planetesimals
\end{keywords}

\vspace{5mm}

\horizontalLine

\noindent \textbf{Contents}

\presecContentsEntry{sec: learningObjectives}
\presecContentsEntry{presec: glossary}
\presecContentsEntry{presec: nomenclature}
\secContentsEntry{sec:introduction}
\secContentsEntry{sec:steps}
\subsecContentsEntry{sec:birth}
\subsecContentsEntry{sec:planetesimals}
\subsecContentsEntry{sec:embryos}
\subsecContentsEntry{sec:migration}
\subsecContentsEntry{sec:giants}
\subsecContentsEntry{sec:instability}
\subsecContentsEntry{sec:rocky}
\subsecContentsEntry{sec:water}
\subsecContentsEntry{sec:moon}
\secContentsEntry{sec:middleage}
\subsecContentsEntry{subsec:asteroids}
\subsecContentsEntry{subsec:milankovitch}
\subsecContentsEntry{sec:demise}
\secContentsEntry{sec:conclusions}

\horizontalLine


\section*{Learning objectives}
\label{sec: learningObjectives}


\noindent By the end of this chapter, you should understand:

\vspace{1mm}

\begin{itemize}
    \item The large-scale orbital structure of the Solar System
    \item The key phases in planetary formation
    \item The events that shaped our Solar System 
    \item The main uncertainties in our understanding of Solar System formation
    \item The present-day Solar System and its evolution on million-year timescales
    \item How and when the Solar System will end
\end{itemize}


\begin{glossary}[Glossary]
\label{presec: glossary}

\term{Accretion}. The growth of planets by the accumulation of smaller bodies.  For example, `planetesimal accretion' refers to growth through collisions with planetesimals. \\
\term{Birth cluster (of stars)}. The birthplaces of stars -- groupings of a few dozen to a hundred thousand stars that typically last for a few million years before dispersing.\\
\term{Differentiation.} A process by which a planet's interior separates into different parts, with iron and {\em sideorophile} elements in the core, rock and {\em lithophile} elements in the mantle and crust, and {\em atmophile} elements in the atmosphere.\\ 
\term{Dynamical instability}. A situation where the orbits of two or more planets cross, such that they can come close to each other.  Dynamical instabilities usually lead to collisions (in the case of rocky planets close to the Sun) or gravitational scattering and ejection from the Solar System (for giant planets far from the Sun).\\
\term{Eccentricity}. The degree of non-circularity of a planet's orbit. The eccentricity is zero for a circular orbit, between zero and 1 for an elliptical orbit, and above 1 for an unbound, hyperbolic orbit.\\
\term{Exoplanets} Planets orbiting stars other than the Sun.  As of 2024, more than 5000 are known.\\
\term{Gas giant planet}. A planet whose mass is primarily made of gas. Gas giants are thought to exist in the range from about half of Saturn's mass ($\sim 50 \mEarth$) up to about 13 Jupiter masses, above which there is some fusion in the object's core and it is instead called a brown dwarf.\\
\term{Inclination}. The angle between a planet's orbit and a reference plane. Among Solar System planets, inclinations are generally no more than a few degrees.\\
\term{Lithophile}. Elements that are `rock-loving', and make up planetary mantles and crusts (as opposed to {\em siderophile} elements, which end up in the cores).\\
\term{Main sequence star.}  A "normal" star that undergoes nuclear fusion of hydrogen into helium in its core (like the present-day Sun).\\
\term{Meteorite}. A rock found on Earth's surface that originated in space.  Most meteorites are fragments of asteroids.\\
\term{Migration.} A process by which a planet's orbit is made smaller (`inward migration') or larger ('outward migration').  Migration is usually driven by gravitational interactions between a planet and the gaseous protoplanetary disk. \\
\term{Milankovitch cycles.} Observed correlations between the Earth's oscillating orbit and spin states and our planet's climate. These cycles have timescales measured in tens to hundreds of thousands of years.\\
\term{Orbital (mean motion) resonance}. When the orbital periods of two neighboring orbits form the ratio of small integers.  For instance, Neptune and Pluto are in 3:2 resonance -- in the time it takes Neptune to complete three orbits around the Sun, Pluto has completed exactly three.  This has a stabilizing effect on Pluto's orbit.\\
\term{Orbital (secular) resonance}.  When the precession rates of two objects match up -- meaning that the direction of their perihelia shift around the Sun at the same rate -- they become strongly gravitationally coupled, and often unstable.  For example, the inner edge of the asteroid belt is marked by a secular resonance with Saturn.\\
\term{Pebble}.  A mm- to cm-sized object that drifts through the gaseous disk.  The accretion of pebbles is a very efficient growth mechanism for giant planet cores.\\
\term{Planetesimal}. Often considered the main building blocks of planets, planetesimals are thought to have typically been 10-100 km in size.\\
\term{Planetary embryo (or core)}.  A massive object that forms from planetesimals.  Embryos are thought to have been Moon- to Mars-sized in the rocky planet zone, and 5-20 Earth masses in the giant planet region. \\
\term{Protoplanetary (or `planet-forming') disk}.  Made of 99\% gas and 1\% dust, disks are found around virtually all young stars.\\
\term{Semimajor axis}. The average orbital distance of an elliptical orbit. \\
\term{Siderophile}.  This describes elements that are `iron-loving' and, when a planet is strongly heated (for example during giant impacts), sink to the core.\\
\term{Streaming instability}. An instability that acts to concentrate dust and pebbles relative to the gas, to promote the formation of planetesimals.\\
\term{Super-Earth}.  A class of exoplanet with a size between that of Earth and Neptune. Super-Earths are found around roughly half of all main sequence stars.

\end{glossary}


\begin{glossary}[Nomenclature]\label{presec: nomenclature}
 
\begin{tabular}{@{}lp{34pc}@{}}
au & Astronomical Unit (the Earth-Sun distance, or $1.496 \times 10^{8}$~km)\\
\hspace{-0.1cm}$\mEarth$ & Earth-mass ($=5.972 \times 10^{24}$~kg)\\
NC & non-carbonaceous (meteorite)\\
CC & carbonaceous (meteorite)\\

\end{tabular}
\end{glossary}

\newpage

\section{Introduction -- Welcome to the Solar System}\label{sec:introduction}

\begin{figure}[t]
\centering
\includegraphics[width=0.9\textwidth]{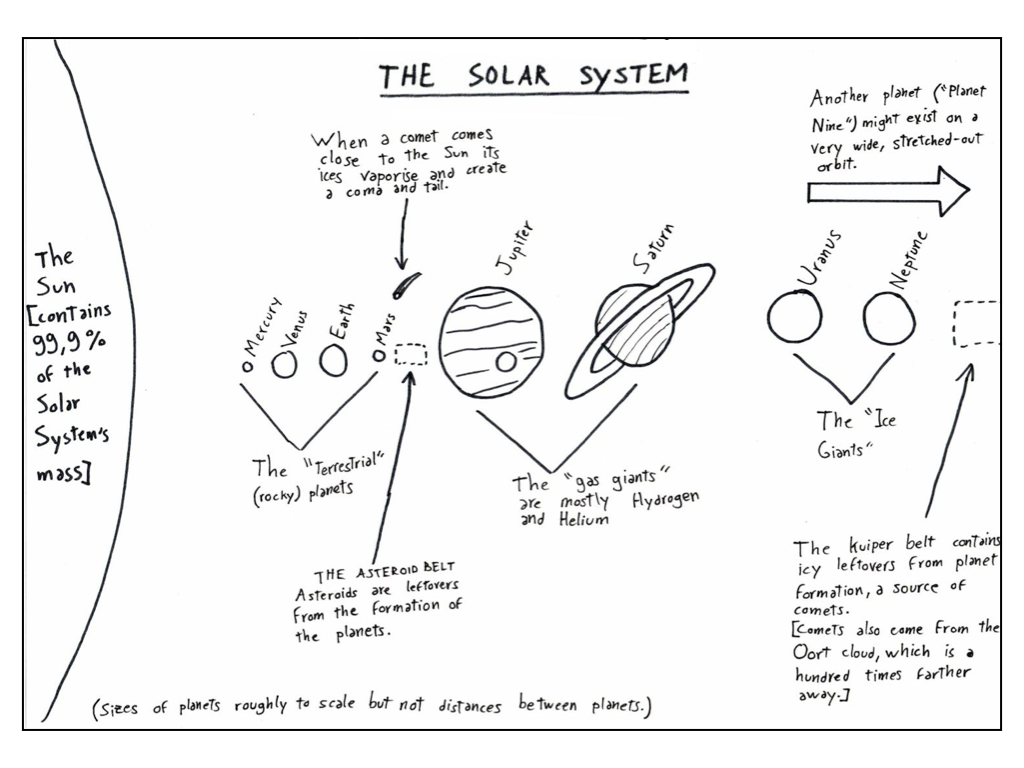}
\caption{Cartoon of the orbital structure of the present-day Solar System.  Credit: Owen Raymond, from {\em Black Holes, Stars, Earth and Mars} (Raymond 2020).}
\label{fig:SS_overview}
\end{figure}

Let's take inventory of our home planetary system (Fig.~\ref{fig:SS_overview}).  The Solar System contains four small rocky planets close to the Sun -- Mercury, Venus, Earth and Mars -- and four giant planets on wider, colder orbits, including two gas giants (Jupiter and Saturn) and two ice giants (Uranus and Neptune).  The rocky (terrestrial) planets add up to a little less than 2 Earth masses ($\mEarth$), whereas the gas giants' masses are each measured in hundreds of Earth masses (Jupiter is $318 \mEarth$ and Saturn is 96 $\mEarth$), and the ice giants are each about 15 Earth masses.  There are also populations of rocky and icy objects (asteroids and comets), which contain a lot of bodies but add up to very little mass.

The orbits of the planets are not perfect circles, but they nonetheless close to circular. Their orbital eccentricities are less than 10\% for all planets apart from Mercury. Their orbital inclinations are less than a few degrees in all cases as well.  This near-circular, near-coplanar configuration is strong circumstantial evidence that the planets formed from a thin disk orbiting the young Sun, as proposed by Laplace.  In contrast, the asteroids' orbits are {\em excited}, with orbital eccentricities from zero to more than 30\% and inclinations from zero to more than $20\deg$.  These small bodies' orbits must have been perturbed somehow from their initial circular, coplanar configuration.

This chapter will tell the Solar System's story, from the point of view of its formation and the planets' orbital evolution. It will start with the birth of the Sun in a cluster of stars, discuss the different steps in the growth of the planets, explain models for the processes that are thought to have been responsible for shaping our system, and end with the death of the Solar System about 100 billion years in the future.  The topics that will be covered include the Sun's birth cluster, the formation of planetesimals and planetary embryos, migration and growth of giant planets, the giant planet instability, growth of the terrestrial (rocky) planets, the Moon-forming impact, the Solar System's middle age, and, finally, the future of the Solar System and its eventual demise.  

The study of our origins is inherently interdisciplinary.  We will intertwine advances from studies of meteorites, observations and models of protoplanetary disks, dynamical (computer) simulations, and studies of planets orbiting other stars (exoplanets).  Each of these fields of study is vital in creating a complete picture of the origins of our system.  It is worth keeping in mind that the story is not fully written, and that there are many areas in which uncertainties remain or are incomplete.

\begin{figure}[t]
\centering
\includegraphics[width=0.8\textwidth]{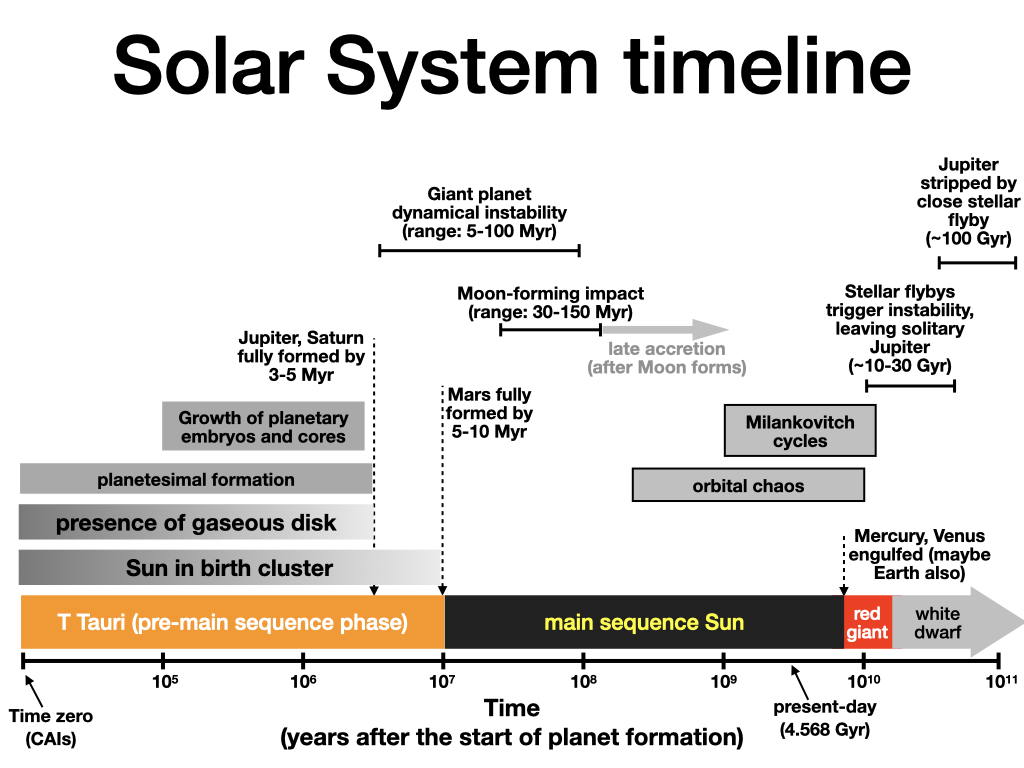}
\caption{A rough timeline of the key events in Solar System history.  Time zero represents the start of planet formation, generally dated using CAIs (Calcium-Aluminum-rich Inclusions, the oldest parts of primitive meteorites).  The Sun's evolution is shown at the bottom. }
\label{fig:timeline}
\end{figure}

\begin{figure}[t]
\centering
\includegraphics[width=0.8\textwidth]{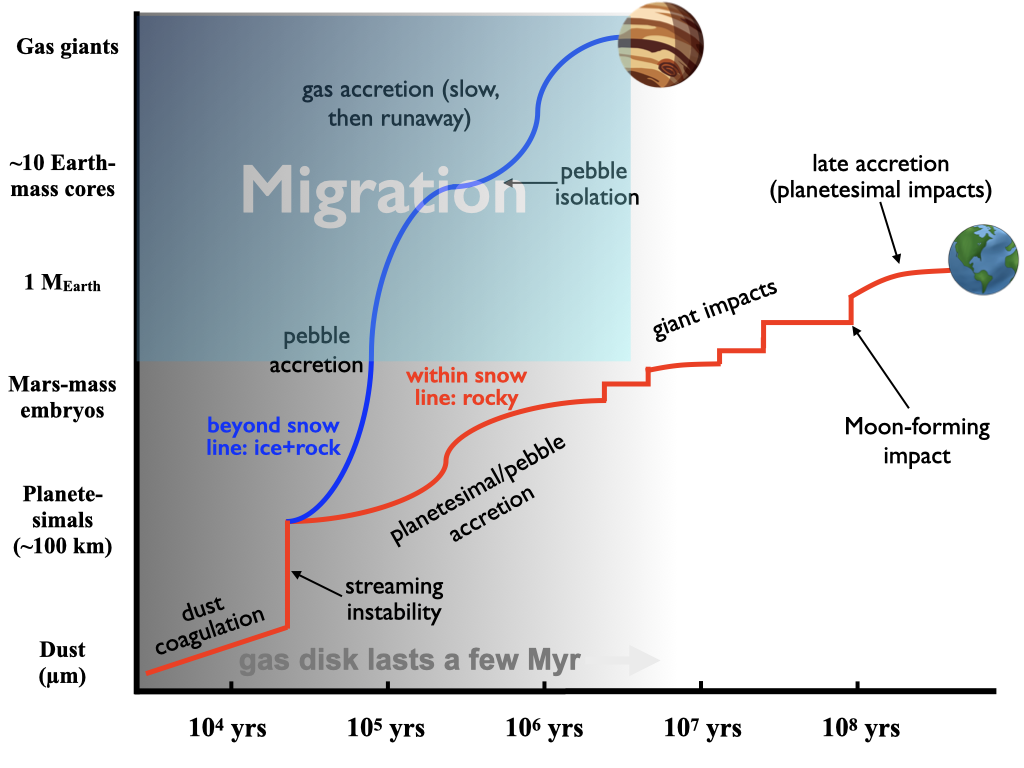}
\caption{Schematic illustration of the key processes involved in planetary formation. The two growth tracks apply to current models for Jupiter and Earth.  Adapted from \cite{raymond22b}. }
\label{fig:mechanisms}
\end{figure}

\section{Key steps in Solar System formation}\label{sec:steps}

Over the past several decades, planetary scientists and astronomers have put together the approximate sequence of events that led to the formation of our Solar System (see Fig.~\ref{fig:timeline}). This section will present an overview of these events. A first step is to understand the setting, starting with the birth cluster of stars in which the Sun was born (\S~\ref{sec:birth}) and the gas-dominated disk that orbited the young Sun.  Next, we will examine the growth of the planets, which started from micron-sized dust grains and proceeded in a number of phases (see Fig.~\ref{fig:mechanisms}).  The first macroscopic objects are thought to have been 10-100 km-scale {\em planetesimals}, which formed directly from the clumping of huge swarms of large dust grains called {\em pebbles} (\S~\ref{sec:planetesimals}).  Planetesimals continued to grow by colliding with other planetesimals as well as eating drifting pebbles, until they reached a size-scale called {\em planetary embryos,} which were roughly the size of the Moon or Mars (about 0.01-0.1 Earth-masses, or $\mEarth$) in the rocky planet region and $5-20\mEarth$ in the giant planet region.  Massive embryos interacted strongly with the gas disk, triggering {\em orbital migration}, the shrinking (or, in some cases, expansion) of their orbits (\S~\ref{sec:migration}).  Massive embryos were essentially giant planet cores, which could -- if they formed fast enough -- gravitationally capture gas to become gas giants (\S~\ref{sec:giants}).  The Solar System's giant planets must have formed during the few million-year lifetime of the Sun's gas disk.  There is strong circumstantial evidence that the orbits of the giant planets become dynamically unstable shortly thereafter, possibly ejecting an extra ice giant (\S~\ref{sec:instability}).  Meanwhile, the rocky planets' formation took longer to complete, ending in a phase of giant impacts between planetary embryos, although the exact processes that shaped their present-day orbits remain a subject of debate (\S~\ref{sec:rocky}).  The final phase of Earth's growth involved a final giant impact with another planetary embryo, the debris from which spawned our Moon (\S~\ref{sec:moon}).  This marked the end of the Solar System's formation.

\subsection{The Sun's birth cluster}\label{sec:birth}

Let's rewind the clock by 4.6 billion years, bringing us to about 9 billion years after the Big Bang.  The curtain lifts on a giant gas cloud orbiting within the Galaxy that is roughly ten light years across and has a total mass of a few hundred to a few thousand Suns. Within the cloud, clumps of gas start to collapse under their own gravity. The centers of clumps reach high enough densities to trigger nuclear fusion and become stars. 

This is how we think the Sun formed; not alone, but in a cluster with perhaps 1000 other stars.  But that cluster is long gone, so how do we know how many stars there were?

There are two pieces of evidence that we can use~\citep{adams10}. First, meteorites contain the decay products of Aluminum-26, which has a half-life of 717,000 years. Only very massive stars make Aluminum-26; `normal' stars like the Sun do not. The Solar System's Aluminum-26 was injected into the Sun's planet-forming disk by a nearby massive star, perhaps a supernova. Stars massive enough to produce Aluminum-26 are rare: statistically-speaking there is only one massive star for every few hundred lower-mass, `normal' stars. So the Sun's birth cluster must have had at least that many. Second, very massive clusters form up to a hundred thousand stars. In a massive cluster, any given star passes pretty close to other stars during its infancy. Very close passages destabilize the orbits of planets or comets far from their stars. Yet there is no sign of a destabilization in the orbits of the ice giants or in the Kuiper belt.  That puts an upper limit on the number of stars in the Sun's birth cluster. 

The Sun's birth cluster must therefore have contained between a few hundred and a few thousand stars. Too few star in the Sun's birth cluster would mean no Aluminum-26. Too many stars and the outer Solar System wouldn't have survived.

Star-forming clusters only last for a few to ten million years. Afterwards, the remaining gas evaporates away into interstellar space. The gas' gravity is what holds clusters together, so with the gas gone, stars (including the young Sun) just drifted apart in the Galaxy.

\subsection{From dust to planetesimals}\label{sec:planetesimals}

Virtually all young stars are seen to have disks around them~\citep{williams11}. These protoplanetary disks contain up to a few percent of the mass of the central star.  They are 99\% hydrogen and helium gas, and only 1\% dust. Yet it is that dust that formed all of the rocky planets, including Earth (which is really just a big rock). When the Sun formed, those ever-so-important solids were just tiny dust grains. They were helpless, blown along with the gas in the disk in orbit around the young Sun. When one dust grain happened to run into a neighboring dust grain, they usually stuck together. The collision speeds were so slow that the dust grains could gently fuse together into long fractal-shaped dust bunnies.

Above a certain size, collisions between dust grains no longer lead to growth. At about the same size scale, large dust grains gain inertia and are no longer just `blowing in the wind,' so to speak. Rather, large grains orbit slightly slower than the gas in the disk and thus feel a headwind that causes them to lose orbital energy and spiral inward~\citep{birnstiel16}. Dust grains that drift through the gas in this way are called pebbles, and are usually about the size of a grain of sand (about 1 mm).

Any individual pebble should drift inward through the disk, and fast~\citep{birnstiel16}. However, if enough drifting pebbles get together, they can clump into much larger objects that don't drift so fast. The required concentration of pebbles is only a factor of a few higher than the background level of dust.  When this happens, a concentration of pebbles acts collectively to slow down the gas. This allows more pebbles to enter the fray, which further slows down the gas and allows even more pebbles to join in. This effect is called the streaming instability. Through the streaming instability, grains of sand-sized pebbles become so concentrated that they clump directly into mountain-sized planetesimals~\citep{johansen14}.

Planetesimals are the building blocks of the planets. They are typically about the size of a big city (10-100 km across). The asteroids and comets are leftover planetesimals -- or fragments of planetesimals -- that didn't make it into planets.  It used to be thought that planetesimals formed in a broad disk, mirroring the distribution of the gaseous disk.  However, images of disks around young stars show that dust is not usually smoothly distributed. Instead, large dust grains (pebbles) tend to concentrate in rings. It's possible that the streaming instability is operating and forming planetesimals within the dust concentrations but not elsewhere. This would form rings of planetesimals, rather than broad disks.  Current models suggest that drifting pebbles may often become concentrated in specific locations within the disk where the gas pressure allows them to gather easily, perhaps where the temperature matches the condensation temperatures of important molecules (like water). 

Planetesimals have the same compositions as their constituent pebbles. Close to the Sun, where it's hot, they're made of iron and rock. But farther away from the Sun, where it is cooler, more volatile molecules exist as solids. The most important is water, which in the very low-pressure environment of the disk, condenses at a temperature of about 170 Kelvin (or -100 degrees Celsius). Beyond the `snow line' water exists as solid ice and any planetesimals that form are about 50-50 ice-rock mixtures (with a little iron included in the `rocky' part). At even colder temperatures other ices exist, like carbon dioxide, ammonia, and carbon monoxide, but they are not abundant enough to have a big impact on planetesimal compositions.

Meteorites can help us understand where and when planetesimals formed.  Most meteorites are small pieces of asteroids. For instance, iron meteorites are pieces of the cores of asteroids, whereas some stony meteorites are pieces of the outer layers of similar asteroids.  Chondrites are a special class of meteorites that appear never to have been part of a very large object, and so their compositions are relatively unaltered. 

\begin{figure}[t]
\centering
\includegraphics[width=0.5\textwidth]{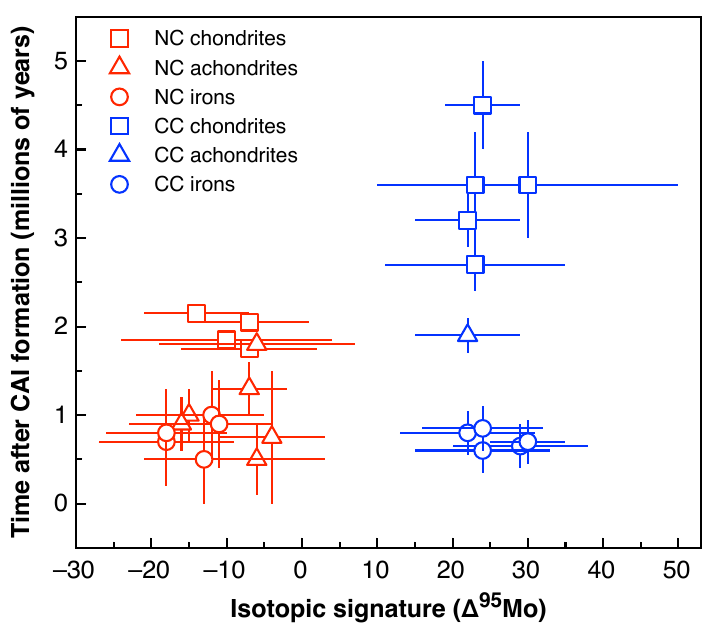}
\caption{The meteorite isotopic dichotomy.  The two-colored groups represent the two classes of  (isotopically-distinct) meteorites: carbonaceous (CC; in blue) and non-carbonaceous (NC; in red).  The parent bodies of the two classes have overlapping age distributions, but different molybdenum isotopic signatures.  Adapted from \cite{kleine20}.}
\label{fig:dichotomy}
\end{figure}

Meteorite science took a giant leap forward when it was discovered that there are really only two fundamental classes of meteorites (see Fig.~\ref{fig:dichotomy}), called carbonaceous (or CC) and non-carbonaceous (or NC). This `dichotomy' is seen in a number of different elements, and it indicates that there were two distinct `flavors' of meteorites forming within the Sun's protoplanetary disk~\citep{kleine20}.  The ages of the two classes overlap, which indicates that the differences between the two classes do not come from a difference in time.  Rather, the differences must instead come from spatial variations.  

Carbonaceous meteorites are associated with objects in the outer parts of the asteroid belt, and non-carbonaceous meteorites with the inner asteroid belt. Most (but not all) carbonaceous meteorites have more water than non-carbonaceous ones. This points toward planetesimals in the outer Solar System being CC-like and in the inner Solar System NC-like.

The planets must have grown from planetesimals that were similar to each of these types of meteorites. Earth's isotopic composition is close to the NC group, but leaning slightly toward the CCs. This leads to the current thinking that Earth grew mostly from (mostly dry) NC meteorite-like planetesimals with just a sprinkling of CC planetesimals (which would have delivered a portion of Earth's water).

We would like to know exactly where and when the Solar System's planetesimals formed.  This is one of the most active areas of planetary science, with a plethora of new models.

\subsection{From planetesimals to planetary embryos} \label{sec:embryos}

After 100 km-scale planetesimals form from concentrations of drifting pebbles, they continue to grow in two ways~\citep{johansen17}: by colliding with other planetesimals (planetesimal accretion) or by collisions with additional drifting pebbles (pebble accretion). 

Collisions between planetesimals often do not lead to net growth, because planetesimals are surprisingly weak. When two crash together they often break apart. After high-speed collisions, the resulting merger can even be smaller than the original objects.  However, the most massive planetesimals can continue to grow rapidly by accreting small planetesimals, even as those smaller ones are often ground into dust. This process appears to have been efficient in the rocky parts of the planet-forming disk, producing `planetary embryos' with masses of a few percent of Earth's mass (between the masses of the Moon and Mars) within just a few hundred thousand years.

Planetesimal accretion fails to produce large planetary embryos farther from the Sun, because above a certain mass the strong gravity of growing cores tends to scatter away planetesimals.  However, pebble accretion has been shown to explain the growth of the cores of the giant planets, growing large planetesimals up to 10-20 Earth masses within less than a million years under reasonable conditions. This fast growth is needed because gaseous protoplanetary disks only last for a few million years before dissipating~\citep{williams11}; this sets an upper limit for the timescale of gas giant formation (although rocky material can persist after the gas dissipates).  Pebble accretion is also self-limiting, as it shuts itself off at a certain mass (of about 20 Earth masses for a Jupiter-like core), by creating a pressure bump that blocks the inward drift of additional pebbles.

Why were embryos so much bigger farther from the Sun? Probably because pebble accretion was much more efficient. Past the snow line, pebbles were cm-sized agglomerations of dust held together by ice. But in the hotter regions close to the Sun, the ice vaporized and the remaining large dust grains were much smaller, only mms in size. Accretion is exponentially faster for larger pebbles. 

\subsection{Orbital migration}\label{sec:migration}

\begin{figure}
\centering
\includegraphics[width=0.3\textwidth]{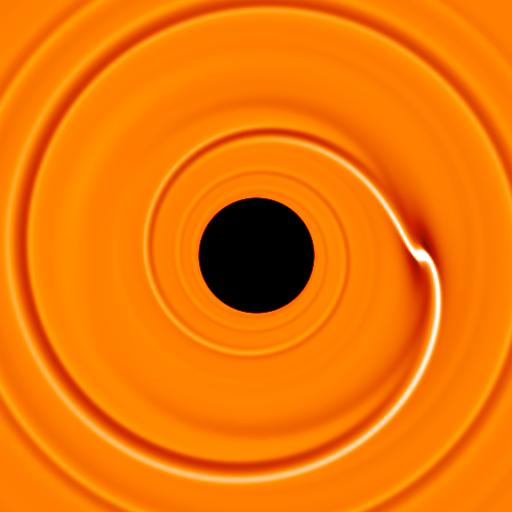} 
\,
\includegraphics[width=0.3\textwidth]{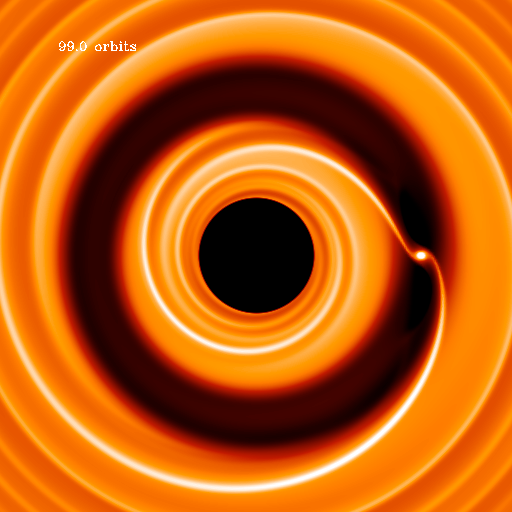}
\caption{Two snapshots of a hydrodynamical simulation of a growing planet at 1 au from its host star interacting with its gaseous nascent disk. In the left panel, a $10 \mEarth$ core has excited density waves in the gas disk -- colors represent perturbations to the gas density.  In the right panel, the core has grown into a Jupiter-mass gas giant and carved a gap in the disk. Credit: Frederic Masset.}
\label{fig:migration}
\end{figure}

Massive planetary `embryos' -- which may grow into the cores of the giant planets -- are generally several times more massive than Earth. A massive embryo (or `core') perturbs the density of the gaseous disk~\citep{kley12}. Perturbations consist of both spiral density waves both interior and exterior to the planet's orbit, and changes to the gas density within the zone that shares the same orbit as the planet (called the `corotation region'; see Fig.~\ref{fig:migration}).  These density perturbations torque the embryo's orbit, causing it to either gain orbital energy and grow, or, more commonly, to lose orbital energy and shrink.  This is called orbital migration.  

Migration is thought to be a fundamental process in planet formation~\citep{raymond22b}.  It is likely to have played a key role in the formation of close-in `super-Earths', planets between the size of Earth and Neptune on close orbits, which exist around up to half of all stars in the Galaxy~\citep{winn15}. The `smoking gun' of migration is when many planets in a given system are locked in orbital resonance, such that pairs of neighboring planets have ratios of their orbital periods that are small integers (for instance, if two planets are in 3:2 resonance, the inner one completes 3 orbits in exactly the same time it takes the outer one to complete 2 orbits.)  Some of the most captivating known exoplanet systems follow this pattern -- the TRAPPIST-1 system is a particularly strong example, with a seven-planet chain of resonances that was almost certainly shaped by orbital migration.  

Migration is faster for higher-mass planets (up to a point; it slows down for gas giants, as we will see below). Below about a Mars-mass ($\sim 0.1 \mEarth$), migration is too slow to matter. But a 10 Earth-mass core migrates on a timescale of 10-100 thousand years -- this is extremely fast given that the gas disks last several millions of years. Migration is not always smooth. If disks are turbulent and have clumps of higher- and lower-density, then there can be a random (`stochastic') element to migration. 

As planets grow more massive, they transition to a different, slower mode of migration~\citep{kley12}. The transition is marked by a planet opening a gap within the disk, after which the planet becomes locked to the disk. Instead of migrating through the gas, it migrates along with the gas as the disk itself evolves. Disks evolve slowly, and most of the gas ends up draining onto the star~\citep{armitage11}. Likewise, a giant planet that has cleared a gap slowly migrates inward (usually). The planet's migration will usually keep going until it either reaches the inner edge of the disk or the disk dissipates.

There are other, fundamentally different types of migration that are not driven by interactions between growing planets and their star's gaseous disks. These all take places after the disk is gone. The two most important are planetesimal-driven migration (when a planet interacts with a disk of planetesimals; this will become important in \S~\ref{sec:instability}) and tidal migration (driven by tidal dissipation within a planet or star; relevant for close-in exoplanets).

\subsection{Growth and migration of gas giants}\label{sec:giants}

Jupiter and Saturn are `gas giants' because they are mostly made of hydrogen and helium gas, with solid cores of about 20 Earth masses. The cores of the ice giants are actually similar in mass to those of Jupiter and Saturn.  The difference is that the gas giants contain most of their mass in gas, whereas the ice giants only have a thin layer of gas on top. 

The most widely-accepted model for the growth of gas giants is the bottom-up, `core accretion' model. In simple terms, a large core grows, then gas is gravitationally captured from the disk and piled on top of the large core. It's easy to imagine that the difference between the gas- and ice giants is related to how fast their cores grew. Fast-growing cores accumulated all the gas they could but slow-growing cores ran out of time, only gathering a small amount of gas before the gas disk itself dissipated. 

As discussed in \S~\ref{sec:embryos}, large cores can form efficiently from pebble accretion onto the largest planetesimals that formed early beyond the snow line~\citep{johansen17}.  Jupiter and Saturn became gas giants by gravitationally capturing gas directly from the disk. The process of gas capture is quite slow at first, because it is limited by how quickly a planet's early gaseous envelope can cool and contract, to make space for more gas.  Once a critical threshold is reached, gas accretion becomes very fast; this threshold is crossed when the mass of the captured atmosphere is about the same as the mass of the core (see Jupiter growth track in Fig.~\ref{fig:mechanisms}). 

Orbital migration plays a key part in giant planet formation, during multiple epochs~\citep{kley12}. First, giant planet cores migrate inward, and fast.  Second, the fully-formed gas giants also migrated, but in a different mode, but probably also inward.  Quite simply, orbital migration is unavoidable and can't be ignored.

This leads to the question: in the face of inward migration, why is Jupiter so far from the Sun?  There are (at least) two possible solutions.  The first suggests that migration really did shrink down the Solar System~\citep{johansen17}.  Jupiter's core really did form at 10 or 20 au and migrated inward as it grew. Saturn and the ice giants formed even farther away (or later) and also migrated. This model can provide a match between models and the present-day Solar System, but it leaves us with another big question: what happened to everything that formed closer than Jupiter's core?  Most of the material interior to Jupiter's core's orbit would have survived Jupiter's migration, so why was there so little of it?

A second possible explanation for why Jupiter is so far from the Sun is that migration was not directed inward or as fast as we might think.  Jupiter's core's migration may have been slowed by effects related to gas heating during pebble accretion, or because the viscosity of disks was lower than generally thought, causing much slower movement of the gas. Alternately, even though a lone gas giant almost always migrates inward, this is not always true for two gas giants together in the same disk~\citep{kley12}. Depending on the properties of the Sun's disk and how fast Saturn grew relative to Jupiter, the gas giants may have either migrated inward together, remained on roughly fixed orbits, or even migrated outward~\citep[this is the basis of the Grand Tack model of][]{walsh11}.  This would have happened after Saturn caught up to Jupiter and the two planets shared a common gap.

Meanwhile, after the gas giants were fully-formed (or at least Jupiter was), the ice giants were probably large cores in the outer Solar System. They must have tried to migrate inward, but they were stopped by Jupiter and Saturn. Exactly where and when this happened depends on the gas giants' migration history.

Regardless of their exact migration pathway, the giant planets almost certainly ended up in a chain of orbital resonances (similar to the Trappist-1 system). Saturn and Jupiter were likely in 3:2 or 2:1 resonance.  This orbital configuration set the stage for the dramatic next phase in the dynamical evolution of the giant planets: the giant planet instability.

\subsection{The giant planet instability}\label{sec:instability}

A huge breakthrough in planetary science came from understanding that the giant planets were probably not born on their present-day orbits. This realization was driven in large part by discoveries of giant exoplanets on very unusual orbits -- in particular, gas giants on extremely eccentric orbits and hot Jupiters~\citep{winn15}.  The main mechanisms thought to have been responsible for the shift in the Solar System's giant planets are planetesimal-driven migration and dynamical instability.  

The present-day Solar System beyond Neptune is quite sparse. If you add up the mass in the entire Kuiper belt and scattered disk -- including Pluto, Eris, Gonggong and more -- it only amounts to about a Mars-mass (about 10\% of an Earth-mass). This does not count Planet Nine, a planet of a few to ten Earth-masses that may exist on a wide, ecccentric, stretched-out orbit~\citep[see Fig.~\ref{fig:SS_overview};][]{batygin19}.  

Yet it is hard to imagine the Kuiper belt as being pristine.  Many Kuiper belt objects have eccentric orbits rather than the circular ones one would expect for objects emerging from the protoplanetary disk.  There also exists a population of objects in resonance with Neptune, the largest being Pluto, which is locked in a 2:3 orbital resonance with Neptune such that Pluto completes two orbits for every three of Neptune.  To explain the orbits of the Kuiper belt objects appears to require a source of dynamical excitation -- possibly from gravitational encounters with planets -- as well as the outward migration of Neptune, which could have captured Pluto and the other resonant Kuiper belt objects~\citep{malhotra93}.  

The Kuiper belt may have originally contained far more mass than it does today.  If that were the case, what would have been the effect on the planets? In a pioneering study, \cite{fernandez84} showed that a massive outer planetesimal disk would have caused the planets' orbits to spread out, with Saturn and the ice giants migrating outward and Jupiter migrating inward. Neptune would have been the first planet to interact with outer disk planetesimals, and scatters them in any direction. Planetesimals that are scattered outward come back again, because it takes a series of scattering events to eject a planetesimal into interstellar space. But planetesimals scattered inward often don't return, because they cross Uranus' orbit and start being scattered by Uranus. So, Neptune scatters more planetesimals inward than outward. The back-reaction of this scattering pushes Neptune outward and causes Neptune's orbit to grow -- this is another form of outward migration. The same process is repeated for Uranus and Saturn: both preferentially scatter planetesimals inward and migrate outward. This is reversed for Jupiter. With its massive gravity and no closer-in giant planets, Jupiter tends to scatter planetesimals outward and eject them. This causes Jupiter to migrate inward.

\begin{figure}[t]
\centering
\includegraphics[width=\textwidth]{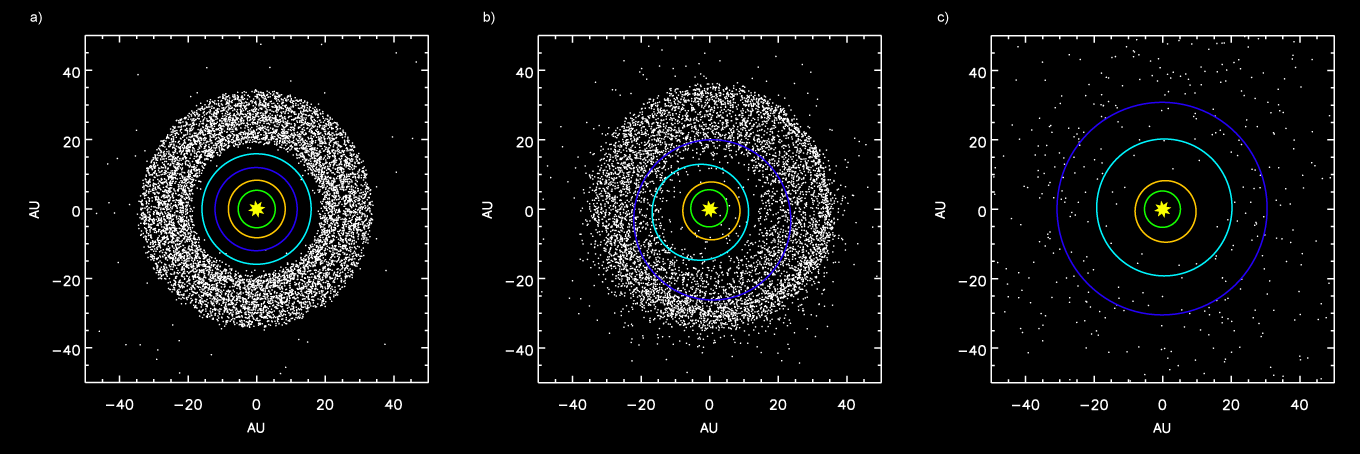}
\caption{Evolution of the Nice model~\citep{nesvorny18}. The solid lines show the orbits of the giant planets, and the white dots represent planetesimals.  The instability is not yet triggered in panel a, is in full swing in panel b, and is complete several million years later in panel c. In this example, the orbits of Uranus and Neptune were reversed during the instability, but no extra ice giant was ejected.  The inner planets and the asteroid belt have orbits that are too compact to be seen on this scale. Adapted from \cite{gomes05}.}
\label{fig:Nice}
\end{figure}

Meanwhile, the discovery of hot Jupiters and giant exoplanet on eccentric orbits spurred the thinking that dynamical instabilities were responsible.  The idea is as follows. When one gas giant forms around a star, it usually has one or two (or many) companion gas giants, and they tend to migrate close to each other (likely into chains of orbital resonances). After the dispersal of the gaseous disk, the planets gravitationally jostle each other and gently stretch out each others' orbits, sometimes to the point that two planets' orbits cross. The planets undergo close gravitational encounters and scatter, exchanging orbital energy. The outcome is that one planet is usually ejected into interstellar space (to roam the galaxy as a free-floating planet). The surviving planets have eccentric orbits, essentially scars of this violent dynamical instability.

The model that put together dynamical instability and planetesimal-driven migration into a coherent picture was the Nice model~\citep[][; pronounced `niece', like the city in France]{gomes05,nesvorny18}. The Nice model assumes that the giant planets formed on a more compact orbital configuration than the present-day one, likely in a chain of orbital resonances. There was an outer planetesimal disk of 20-30 Earth masses (see Fig.~\ref{fig:Nice}). There was likely an extra ice giant, making 5 total giant planets at the start. (The reason for this is that one ice giant is almost always ejected, and the model needs to explain that four giant planets survived.) 

When the instability was triggered, the extra ice giant was ejected into interstellar space after an encounter with Jupiter. Uranus and Neptune interacted with the planetesimal disk and triggered a phase of planetesimal-driven migration. The ice giants migrated outward, and Neptune reached close to the outer edge of the disk, capturing Pluto and its cohort in the process~\citep{malhotra93}.  Most of the rest of the planetesimal disk was ejected by Jupiter (although some planetesimals were implanted into the Oort cloud). The giant planets spread out and reached their current orbits.

The Nice model has been remarkably successful in reproducing several hard-to-explain aspects of the Solar System~\citep{nesvorny18}, including the giant planets' orbits, the orbital distribution of Jupiter's Trojan asteroids (which represent planetesimals from the outer planetesimal disk that were captured during the instability), the irregular satellites of the giant planets (also captured), and the orbital structure of the Kuiper belt and scattered disk (shaped by the giant planets during the instability). The fact that the model can explain all of these different characteristics of the Solar System is why the planetary science community has adopted this idea (now usually called the `giant planet instability' model) as the paradigm for the early evolution of the outer Solar System. 

However, there are two big uncertainties with the model: the instability trigger and timing. The Nice model was originally designed to reproduce the `terminal lunar cataclysm', a perceived spike in the rate of impacts on the Moon about 500-600 million years after the start of planet formation~\citep{bottke17}. The concept of this cataclysm originated with analyses of craters on the Moon, supplemented by samples brought back to Earth by the Apollo missions, from several different sites on the Moon.  Chemical analyses of Moon rocks measured their ages with radioactive dating techniques. Combined with information about the craters where the rocks were collected, analyses calculated the rate of impact bombardment on the Moon throughout its lifetime (the `impact flux').

Several analyses inferred the presence of a delayed spike in the Moon's impact flux. This spike, or cataclysm, corresponded to the time of the giant planet instability in the original Nice model~\citep{nesvorny18}. In recent years, evidence for the terminal cataclysm model has disappeared~\citep{zellner17}. New analyses have shown that the rocks brought back from the Moon suffer from sampling bias. The giant collisions that produced the youngest impact basins on the moon -- especially Imbrium -- threw rocks across the surface of the Moon and contaminated the samples collected elsewhere. The `spike' in impact rates was just a single big impact. Rather than a terminal cataclysm, it is now thought that the impact rate on the Moon declined smoothly. 

From a dynamical point of view, the giant planet instability model has been remarkably successful.  But when did this instability take place?  Analyses of different parts of the Solar System -- including highly-siderophile elements, the existence of a binary Trojan asteroid, and impact reset ages of meteorites -- all agree that the instability must have happened early. To be more precise, the instability could not have taken place later than 100 million years after the start of Solar System formation, and possibly significantly earlier.

The key factor that would determine the timing of the instability is, simply, the instability trigger.  There are three possibilities to date.  First, the instability could have been triggered by the dispersal of the gaseous disk.  Disks do not dissipate smoothly at all orbital distances at the same time~\citep{armitage11}. Rather, high-energy radiation from the central star creates an inner hole in the disk, which then widens in time until the disk is gone. The inner edge of the disk expands in time, and this expansion can push planets along with it, effectively driving outward migration of the planet that is right at the disk edge.  This can act to `squish' planets' orbits closer together and trigger dynamical instability~\citep{liu22}.  A second possibility is that the planets' orbits were simply unstable on their own after the gas disk dissipated.  Finally, the most-studied possibility is that interactions between the giant planets (in their chain of resonances) and the outer planetesimal disk caused the planets' orbits to spread out until they reached an unstable configuration, triggering instability.  The first two triggers would have caused the instability to happen within a few to ten million years after dispersal of the Sun's gaseous disk, whereas the third trigger would have taken 30-60 million years to operate~\citep{ribeiro20}.

After the dynamical instability, the giant planets' orbits would have been basically set, with minimal changes between then and the present day.

\subsection{Formation of the terrestrial (rocky) planets}\label{sec:rocky}
One of the biggest challenges in planetary science is to reproduce the Solar System in a computer.  One of the hardest parts is matching the terrestrial planets~\citep{raymond22b}.  The goal is to match the overall distribution of the planets: the number of planets, their orbits, masses and compositions (and water on Earth). We also want to reproduce the asteroid belt -- to explain why it contains so little mass (less than 1/1000th of an Earth-mass), why the asteroids' orbits are stretched-out (`eccentric'), and why the inner belt contains dry asteroids and the outer belt mostly hydrated ones~\citep{demeo13}.

There is currently no universally agreed-upon story for the formation of the rocky planets. Nonetheless, cosmochemical constraints suggest that Earth's growth was not complete until 50-100 million years after the start of planet formation~\citep[][see also Fig.~\ref{fig:mechanisms}]{kleine17}.  In contrast, Mars' was complete within just 5-10 million years.  We don't have cosmochemical constraints for Venus or Mercury.  However, the gas giants were fully-grown before the dispersal of the gaseous disk at 3-5 million years. 

The foundational model of rocky planet formation is the so-called `classical model'~\citep{raymond22b}. The main assumption of the classical model is that the growth of the giant planets can be considered separately from the growth of the rocky planets. Simulations of the classical model do a decent job of matching Earth and Venus.  However, they systematically fail to match Mars, instead forming planets in Mars' orbital vicinity about the same mass as Earth. This is the `small Mars problem'~\citep{wetherill91}. The small Mars problem exists because, within a disk of planetesimals or planetary embryos, neighboring planets tend to form with similar masses (or sizes). The fact that Mars is so much smaller than Earth means that either some process removed mass from Mars' feeding zone (but not Earth's), or that mass was never there to begin with.  The classical model also fails to match the asteroid belt.  To reproduce the present-day belt, the primordial belt (generally assumed to have started with $1-2 \mEarth$ in the classical model) would need to be depleted by 3 orders of magnitude.  This rarely happens in simulations.

There are currently six models that provide viable solutions to the small Mars problem and can match the inner Solar System in broad strokes. I will provide a brief overview of the key processes at play in each model, as well as each model's potential Achilles heel. 

\vskip .1in
\noindent \textbf{The empty primordial asteroid belt model}

This model proposes that the starting conditions for terrestrial planet formation was a ring of planetesimals between roughly the orbits of Venus and Earth.  This is a viable outcome of the planetesimal formation process, given that rings of pebbles are commonly observed in images of planet-forming disks, and that such concentrations could easily represent the sites of planetesimal formation~\citep{izidoro22}.  In the context of a ring of rocky planetesimals, Earth and Venus grow big within the ring. Mars would have been scattered out of the ring and its growth stunted. The same would have happened to Mercury on the inner side of the ring. Simulations show that the empty asteroid belt model can do a remarkably good job of matching the terrestrial planets' masses, including Mars. 

The model assumes that no planetesimals formed between the orbits of Mars and Jupiter, such that the primordial asteroid belt was indeed devoid of macroscopic bodies.  Nonetheless, asteroids would naturally have been implanted both from the giant planet-forming region (in the form of volatile-rich, C-type asteroids) and from the terrestrial planet-forming region (as more refractory-rich, S-types), providing a very good match to the present-day belt~\citep{raymondnesvorny22}. 

Put together, the empty primordial asteroid belt model provides an elegant scenario, as long as planetesimals really do form in rings.  Indeed, understanding exactly where and when planetesimals formed is the key limitation of this model.

\vskip .1in
\noindent \textbf{A pebble accretion scenario}

The two main ingredients in the pebble accretion model are pebble accretion and orbital migration. The assumption is that planetesimals only formed at a certain location in the inner Solar System, although there were multiple generations~\citep{johansen21}. Each generation of planetesimals formed a little past Mars' orbit. The most massive planetesimals ate other planetesimals and then started to accrete drifting pebbles. As they planetary embryos grew more massive, they started to migrate inward. The next generation of planetesimals again formed an embryo capable of eating pebbles, following in the footsteps of the previous one. So, more massive planetary embryos were farther from the site of planetesimal formation. Of course, Earth is more massive than Venus, yet Venus is closer to the Sun and presumably would have grown earlier and migrated farther. However, Earth underwent a late giant impact after the gas disk dissipated, which led to the formation of the Moon~\citep{kleine20}. The pebble accretion model works if there was no `Earth' at this stage but rather two half-Earths that would later collide to form the Earth. Also, Mercury must have formed in a different way, perhaps from iron-rich pebbles that only existed much closer to the Sun.

The uncertainties in this model are related to whether the pebble accretion model can match the asteroid belt and whether it fits with data from meteorites. A potential Achilles heel of this model is whether planetesimals truly form in a given location, as well as confronting it with meteorite measurements. Both are areas of vigorous study right now.

\vskip .1in
\noindent \textbf{The convergent migration model}

Migration is not always directed inward; in some places, migration can be outward. If there are zones of outward- and inward migration in the same disk, there can be regions toward which planetary embryos migrate. These are called convergence zones. The convergent migration model~\citep{broz21} proposes that, within the Sun's gaseous disk, there was a convergence zone located right around Earth's orbit. As planetary embryos grew, the most massive migrated into the center of the convergence zone, where they could grow further. The small ones stayed on the fringes. This migration effectively created a sort of ring of embryos that was broadly similar to the ring of planetesimals in the empty primordial asteroid belt model. This explains why the most massive terrestrial planets -- Earth and Venus -- are located close together, whereas Mercury and Mars are on the outskirts.

While successful in matching the terrestrial planets, it remains to be seen whether the convergent migration model can explain the asteroid belt. A potential Achilles heel is simply whether convergent migration zones actually exist in the right part of protoplanetary disks.

\vskip .1in
\noindent \textbf{The Grand Tack scenario}

In the Grand Tack model, Jupiter's migration cleared out the asteroid belt and Mars' feeding zone, in two parts. First, Jupiter migrated inward on its own. Then Saturn migrated inward and caught up to Jupiter. Then the two planets `tacked' and migrated back outward~\citep{walsh11}. Jupiter could not have migrated too close to the Sun because it would have prevented the formation of the rocky planets. Jupiter's turnaround point must have been at about 1.5 or 2 astronomical units. Jupiter's inward migration compressed the disk of planetary embryos and planetesimals into a narrow ring, and then it migrated back outward and let the terrestrial planets grow in a way that was similar to the empty primordial asteroid belt model.

The Grand Tack model is robust but does have a potential Achilles heel: the outward migration mechanism itself. It has been shown that Jupiter and Saturn should indeed migrate outward but only if certain conditions are met. The key condition is that the Jupiter-to-Saturn mass ratio must be between about 2 and 4. Today, that ratio is about 3.3, but while the gas giants were embedded in the gaseous disk, they were simultaneously migrating and growing. It remains to be seen whether long-range migration could have been maintained in the face of realistic gas accretion.

\vskip .1in
\noindent \textbf{The Early Instability model}

The Early Instability model connects the dots between the outer and inner Solar System~\citep{clement18}. Its central ingredient is the dynamical instability that is thought to have taken place among the giant planets~\citep{nesvorny18}. As described above, the dynamical instability was a violent epoch in the Solar System's history.  If it happened early (as is quite likely), the gravitational scattering of the giant planets would have strongly depleted the primordial asteroid belt, as well as Mars' feeding zone.  However, Earth's feeding zone and the inner regions of the system would not have been strongly affected~\citep{clement18}.  

The Early Instability model matches the terrestrial planets and asteroid belt by making use of an event that almost certainly took place, the giant planet instability. What is compelling is that simulations of the Early Instability model best reproduce the inner Solar System when they also match the outer Solar System. Its Achilles heel is knowing exactly when the giant planet instability actually took place. If the instability took place later than about 20 million years after the start of Solar System formation, it would be too late to provide a solution to the small Mars problem.

\vskip .1in
\noindent \textbf{The Jupiter-Saturn resonant chaotic excitation model}

Emerging from the gaseous disk, Jupiter and Saturn's orbits may have been in 2:1 resonance.  In that case, their orbital evolution may have been chaotic, with strong secular resonances chaotically jumping across the inner Solar System.  Simulations show that excitation from these chaotically-jumping resonances would have destabilized objects in the asteroid belt and also in Mars' feeding zone, causing them to be cleared out~\citep{lykawka23}. The model can provide a good match to the terrestrial planets and asteroid belt.

The main uncertainty with the Jupiter-Saturn resonant excitation model is the timing of events.  The chaotic excitation mechanism would have needed at least a few million years to operate to deplete the asteroid belt and Mars' feeding zone.  This means that the giant planet instability could not have taken place too early for this model to remain viable.

\vskip .1in

Each of these six models can reproduce the rocky planets and the inner Solar System, at least in broad strokes, when certain assumptions are made. Yet in all cases, the final phases of growth involved a series of giant impacts between planetary embryos (see Earth's rough growth track in Fig.~\ref{fig:mechanisms}).  These models are not mutually exclusive, and it is possible that more than one took place -- for instance, the giant planet instability may have taken place early {\em and} the rocky planets' building blocks may have been concentrated in a narrow ring.  

\subsection{The origin of Earth's water}\label{sec:water}

Water is an essential part of our planet.  Yet, despite being covered in oceans, Earth is not a water-rich planet.  Its overall water budget (by mass) is only about 1 part in 1000. Earth's composition was set during its formation~\citep{krijt23}.  While computer simulations are the tool of choice to show how planetary embryos were assembled into planets, cosmochemical measurements are useful for understanding the origin of Earth's water, which in turn has implications for what source material Earth was formed from~\citep{steller22}.

The Deuterium-to-Hydrogen -- or D/H -- ratio of Earth's water is $1.56 \times 10^{-4}$.  The D/H ratios observed in outgassing from comets are typically higher, whereas the D/H ratio of Jupiter's atmosphere and of the Sun are each 6 times lower.  

The most important comparison sample are meteorites, which are fragments of asteroids fallen to Earth. They represent the building blocks of the planets (at least, the ones that didn't end up in the planets) and so may reflect their compositions.  Enstatite chondrite meteorites, pieces of planetesimals formed in the terrestrial planet-forming region, have been shown to have D/H values that come close to matching Earth's.  Likewise, carbonaceous chondrites, fragments of asteroids from the outer parts of the main belt, also have similar (but not indentical) D/H values to Earth's.  Enstatite chondrites are non-carbonaceous (NC in Fig.~\ref{fig:dichotomy}) whereas carbonaceous chondrites are, naturally, carbonaceous (CC in Fig.~\ref{fig:dichotomy}).  Based solely on the D/H ratio, there is no single solution for the relative contribution of CC and NC planetesimals needed to match Earth.

The story becomes clearer when considering isotopes of other volatile elements, nitrogen and zinc.  To match the isotope ratios of all three volatile elements (including hydrogen) requires a roughly 70-30 mixture of water (or hydrogen) from NC (enstatite) and CC (carbonaceous) planetesimals.  Given the much higher concentration of water in CC meteorites relative to NC, this balance happens with a roughly 95-5 mixture of NC and CC planetesimals~\citep{steller22}. This indicates that Earth must have formed predominantly from NC planetesimals that grew in the inner parts of the Sun's protoplanetary disk, but with a 5\% `sprinkling' of CC planetesimals that originated past Jupiter.  This CC contribution was likely in the form of planetesimals scattered inward during Jupiter and Saturn's growth and migration~\citep{raymond22b,raymondnesvorny22}.  Each of the terrestrial planet formation models presented in \S~\ref{sec:rocky} can match Earth's water.

\subsection{The Moon-forming impact}\label{sec:moon}
The Moon is about a quarter the size of Earth and 1/80th the mass. It is similar in composition to Earth's mantle, although it is quite dry. Of particular importance, the Moon has nearly identical Oxygen isotopes to Earth.

The giant impact hypothesis is the leading model for the Moon's formation~\citep{canup21}. In broad strokes, this hypothesis envisions a giant impact between the proto-Earth and another planetary embryo, which would have generated a disk of debris and vapor around the young Earth, from which the Moon coalesced.  

There are several flavors of Moon-forming impacts. The `canonical' Moon-forming impact envisions the collision between the almost-formed Earth (with about 90\% of its current mass) and a Mars-mass planetary embryo. The canonical impact produces a Moon with the right orbit, but most of the Moon's mass comes from the impactor, called Theia. To explain why the Earth and Moon have such similar compositions (especially in Oxygen isotopes), Theia must have grown very close-by to Earth, or at least been made of a similar mix of materials. A more energetic impact would mix the proto-Earth and Theia together such that the Earth and Moon's compositional match comes from the fact they were both made from the same mix. More energetic impacts can involve either planetary embryos that were spinning very fast before the impact, or an impact between roughly equal-mass planetary embryos. Either case can explain the Earth-Moon compositional similarity (at least in broad strokes).

The giant, single-impact is currently favored, but there are plenty of alternate scenarios and uncertainties (Canup et al 2021). For example, the Synestia hypothesis proposes that the impact was so energetic that it generated a giant torus of rock vapor, whose evolution determined the Moon's final properties. Another model suggests that the Moon formed not from a single giant impact but from a series of smaller ones.

The Moon-forming impact is thought to have been Earth's final giant impact. Other, later impacts of planetary embryos would likely have destabilized the Moon from its orbit. The time of the Moon-forming impact therefore measures the end of Earth's main phase of growth (apart from late accretion-- see below and Fig.~\ref{fig:mechanisms}).

The Hafnium-Tungsten radioactive system can be used to measure the timing of the Moon-forming impact~\citep{kleine17}. Hafnium (182-Hf) decays into Tungsten (182-W) with a half-life of 9 million years. Hafnium is lithophile; it `follows the rocks'. But Tungsten is siderophile; it `follows the iron'. The amount of Tungsten in Earth's crust and mantle can be used as a clock to tell us when Earth differentiated. If Earth had differentiated very quickly, long before any Hafnium had the time to decay into Tungsten, all of the Tungsten would have remained in the mantle and crust. If, instead, Earth differentiated very late, long after all of its Hafnium had decayed into Tungsten, then all of Earth's Tungsten would have sunk to the core and none would remain in the mantle or crust.

The current estimate is that the Moon-forming impact took place about 40 to 150 million years after the start of planet formation~\citep{kleine17}. Simulations of terrestrial planet formation can match this timescale, for each of the scenarios discussed in \S 2.7.

After the Moon formed, its orbit relative to Earth was controlled by tidal evolution. The Moon's orbit expanded dramatically as it was tidally pushed away from Earth. The Moon and Earth's spins were also affected. The Moon is still tidally evolving today, moving away from the Earth at a rate of about 3.8 cm per year (about the same rate at which fingernails grow).

The final phase of Earth's growth is called `late accretion' (or, sometimes, the `late veneer').  Late accretion consisted of planetesimals hitting the Earth after the Moon-forming impact; there were no embryo impacts, as those would have triggered another differentiation event.  

We can estimate how much mass was delivered by late accretion using the abundance of siderophile elements in the mantle~\citep{walker09}.  The Moon-forming impact triggered differentiation of Earth, meaning that iron and all siderophile elements were sequestered into the core. The siderophile elements in Earth's mantle and crust today came from planetesimal impacts after the Moon-forming impact. Assuming that the compositions of primitive (chondritic) meteorites are a reasonable proxy for the composition of impactors, measurements indicate that late accretion contained only about the last half of a percent of Earth's mass.  Simulations indicate that the vast majority of late accretion took place within a few hundred million years after the end of planet formation.  

This truly marks the end of Earth's -- and the Solar System's -- formation.

\section{The Solar System's middle age and future}\label{sec:middleage}

Let us now consider the rest of the Solar System's history, namely, its present and future.

The planets, asteroids and comets have survived for four and a half billion years.  But that does not mean that the Solar System is completely static.  The orbits of the planets themselves are in a constant, rhythmic dance; oscillations in the shapes and alignments of Earth's orbit play a key role in the long-term evolution of our climate. The asteroids and comets have evolved, and a small fraction have even collided with the planets (including Earth). 

The Solar System's future will be dynamic (and, eventually, tragic).  Chaos pervades the long-term orbital evolution of the rocky planets, and there is a small chance that they will become unstable in the next few billion years.  Our Sun itself is slowly evolving, and in $\sim 7$ billion years will puff into a red giant, then lose its outer layers to become a white dwarf.  Like an ant in a rainstorm, the surviving planets will be at the mercy of stars flying past the Sun.  

\subsection{Asteroids and comets}\label{subsec:asteroids}

Asteroids and comets are the leftovers from the planets’ formation. They’re like the potato peels that end up on the floor instead of in the mashed potatoes.  

The asteroid belt a broad swath of Solar System real estate that extends from Mars’ orbit out to Jupiter’s.  The vast majority of asteroids are located in the main belt, between about 2.1 and 3.2 au.  The belt contains less than one one-thousandth of Earth’s mass, although there are still more than a million asteroids larger that 1 kilometer in diameter. The belt has a lot of substructure in terms of compositional classes of asteroids and asteroid ‘families’.  Of particular note, the inner parts of the main belt are dominated by relatively dry asteroids and the outer belt by more volatile-rich and water-rich asteroids~\citep{demeo13}.

There are two distinct potential origins stories for the asteroid belt: one of depletion and one of implantation~\citep{raymondnesvorny22}. Older models of terrestrial planet formation such as the classical model and {\em Grand Tack} (\S~\ref{sec:rocky}) assume that planetesimals formed in a broad disk from Venus' orbit out past Jupiter's, with at least $1-2\mEarth$ in the belt.  These models therefore require a depletion of a factor of 1000 or more to explain the belt's present-day low mass.  In contrast, the {\em Empty primordial asteroid belt} model takes the opposite approach, by proposing that no planetesimals ever formed in the belt.  It therefore requires a modest number of asteroids to have been implanted during the planets' growth.  

The asteroid belt is crisscrossed by gaps at orbital resonances with Jupiter and Saturn, called the Kirkwood gaps. An asteroid in a Kirkwood gap is unstable: any objects that enter the gaps are quickly removed. In practice, upon entering a strong resonance such as those that carved the Kirkwood gaps, an asteroid's orbital eccentricity is excited until it crosses the orbit of a planet (or, in some cases, until it collides with the Sun).  The asteroid is then scattered by the planet and leaves the belt.  The asteroid will eventually either be ejected from the Solar System (usually after encountering Jupiter) or collide with one of the rocky planets (after encountering Mars).  The Kirkwood gaps were cleared out early in Solar System history, but small asteroids (or asteroid fragments) are still slowly pushed into unstable resonances by thermal forces such as the Yarkovsky effect~\citep{vok15}.  This is the source of the continued population of impactors on the Earth and rocky planets, including the dinosaur-killing impactor 65 million years ago.  

Asteroids and comets are closely related.  Asteroids are usually thought of as rocky and comets as icy, but that distinction does not always hold true. There are ‘active asteroids’ in the main asteroid belt that exhibit comet-like activity, as well as purely rocky objects on orbits similar to those of comets.  

Comets are generally distinguished from asteroids by their orbits.  Comets may enter the inner Solar System, but only in passing.  They originate in two main reservoirs: the Kuiper belt and Oort cloud~\citep{fraser22}.  The Kuiper belt contains a total of about 0.1 $\mEarth$, although it is likely that the primordial belt contained $\sim 20 \mEarth$ (see Section~\ref{sec:instability}).  The Oort cloud contains a total of $\sim 1 \mEarth$, extending out to the edge of the Solar System at $\sim 100,000$~au.  (Beyond this distance, the Galaxy's tidal field rapidly strips a comet away from the Sun).   Both of these populations of cometary nuclei (the precursors of comets) were put in place during the giant planet instability~\citep{kaib22}.  When the giant planets went unstable, the primordial disk of cometary planetesimals was disrupted.  A large fraction were ejected entirely from the Solar System, but models suggest that 5-10\% were captured in the Oort cloud, with their orbits being randomized in inclination by torques from the Galactic tide and passing stars~\citep{kaib22}.  Meanwhile, only $\sim$0.1\% were trapped in the Kuiper belt, including various sub-populations such as the resonant Kuiper belt objects (such as Pluto), and the scattered disk.

The Kuiper belt is the source of Jupiter-family comets (JFCs), which tend to follow the ecliptic plane~\citep{fraser22}.  JFCs are scattered inward by the giant planets one at a time, following the pattern noticed by \cite{fernandez84} when studying planetesimal-driven migration (see \S~\ref{sec:instability}).  The Oort cloud is the source of long-period comets (LPCs), which may enter the inner Solar System from any direction in the sky. 

Cometary impacts on Earth are far less frequent than asteroidal ones.  

\subsection{Milankovitch Cycles and orbital chaos} \label{subsec:milankovitch}

The planets' orbis are constantly changing their shapes.  Due to mutual gravitational interactions, each planet undergoes oscillations in orbital eccentricity and inclination (although their semimajor axes are basically fixed).  On a timescale of $\sim$100,000 years, Earth's eccentricity varies between almost zero and about 0.06~\citep{quinn91}. Earth's obliquity oscillates between about 22 and 24.5 degrees every 41,000 years, and its spin axis completes a precession cycle every 26,000 years.  

\begin{figure}[t]
\centering
\includegraphics[width=0.6\textwidth]{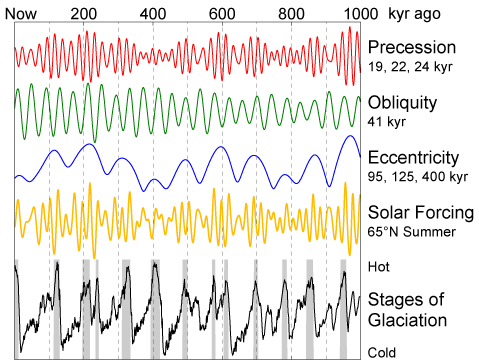}
\caption{Earth's Milankovitch cycles.  The plot simultaneously displays variations in Earth's orbital and spin quantities~\citep{quinn91}, as well as the Earth's mean global temperature (bottom; ice ages are shaded).  Credit: Robert A. Rohlke, for the Global Warming Art project, via Wikimedia Commons.}
\label{fig:milankovitch}
\end{figure}

Milankovitch (1941) proposed that oscillations in Earth's orbit and spin were key drivers of our planet's climate on long timescales.  This was confirmed by empirical measurements~\citep[][; see Fig.~\ref{fig:milankovitch}]{hays76}.  The net effect of an eccentric orbit is to increase the variations in Solar flux (and therefore in surface temperature) throughout the year. Compared with a planet on a circular orbit, a planet on an eccentric orbit passes both closer to the Sun at perihelion and farther from the Sun at aphelion.  Meanwhile, a planet's obliquity determines which latitudes receive the most energy from the Sun.  

Earth's orbital and spin evolution thus plays a central role in the long-term evolution of our planet's climate.  Reconstructions of the climate extending into the distant past have been enabled by simulations of Earth's orbital quantities. The dominant cycle with respect to ice ages is the $\sim$100,000-year oscillation of Earth's eccentricity (see Fig.~\ref{fig:milankovitch}). However, while the variations that govern Milankovitch cycles appear to be regular, their long-term behavior is governed by orbital chaos (at least for the terrestrial planets), with a Lyapunov timescale of $\sim 5$~million years~\citep{laskar94}.  

The orbital evolution of the terrestrial planets has been shown to be chaotic, with a Lyapunov time on the order of 5 million years~\citep{laskar94}.  This is a limiting factor in attempting to align Earth's past orbital evolution with geological samples, to understand our planet's climate earlier than 50-70 million years in the past.   

Looking to the future, the terrestrial planets' chaotic evolution has the potential to trigger an instability among the rocky planets.  In some simulations of the future of the Solar System, Mercury becomes trapped in a secular resonance with Jupiter, in which its orbital precession matches that of Jupiter~\citep{laskar94}.  This resonance will drive up Mercury's eccentricity until it either hits Venus or collides with the Sun (although there is a small chance it could trigger a terrestrial planet-wide instability, possibly resulting in a collision between Earth and another rocky planet).  The probability that Mercury's orbit will be destabilized in the next five billion years is $\sim$1\%~\citep{laskar94}.

\subsection{The future of the Solar System and its eventual demise} \label{sec:demise}

Over the next several billion years, a series of unfortunate events will take place in our Solar System, spanning from the not-so-great to the truly tragic. At the end, our system will be gone: all of the planets will be lost and the Sun will be a solitary white dwarf. 

The first phase involves the over-heating of Earth.  The Sun's luminosity is slowly growing as it evolves along the main sequence.  In 1-2 billion years, the inner edge of the habitable zone -- the ring of orbits inside which a planet can plausibly have liquid water on its surface, if certain conditions are met -- will sweep past Earth's orbit.  The oceans will boil and Earth will no longer be a habitable world.  


\begin{figure}[t]
\centering
\includegraphics[width=0.6\textwidth]{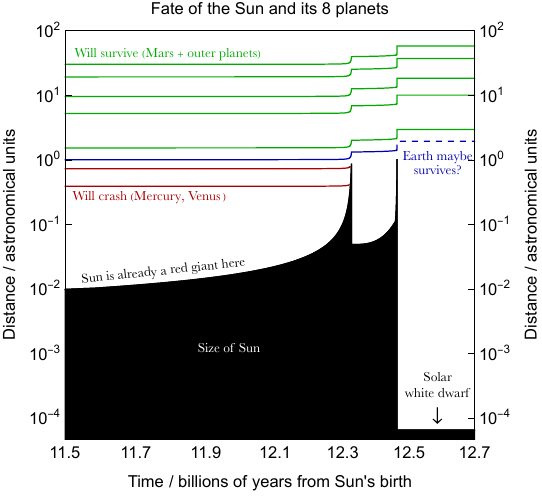}
\caption{Evolution of the Sun and the planets' orbits during the Sun's post-main sequence evolution, starting at the very end of the main sequence phase, just after the Sun becomes a red giant.  The shaded region shows the Sun's radius as a function of time, as it puffs up, then loses its outer layers and is left as a white dwarf.  The planets' orbital radii respond to this evolution in dramatic fashion: Mercury and Venus are engulfed during the red giant phase, whereas the orbits of Mars and the giant planets expand by a factor of $\sim 2$~\citep{veras16}.  Earth may survive or be swallowed by the red giant Sun depending on tidal dissipation parameters. Credit: Dimitri Veras. }
\label{fig:redgiant}
\end{figure}

In $\sim 7$ billion years, the Sun's core will run out of hydrogen and the Sun will puff up into a red giant.  It will engulf Mercury and Venus.  Earth is on the cusp; depending on factors related to tidal dissipation, Earth may either be pushed outward or swallowed by the red giant Sun~\citep{veras16}. During this time, the Sun will be so bright that the habitable zone will extend out to Jupiter and Saturn's orbits.  After about seven hundred million years as a red giant, the Sun's core will again run out of fuel.  Its outer layers will puff off as a planetary nebula, leaving behind the core: a white dwarf.  During the period of mass loss, the planets' orbital radii will expand by almost a factor of two.  

The final phase of the Solar System's evolution will be determined by flybys of other stars~\citep{zink20}.  On a timescale of 10-30 billion years, a star will fly close enough to trigger a dynamical instability among the surviving planets.  This will lead to strong planet-planet scattering and, most likely, the ejection of all planets apart from Jupiter.  On a timescale of $\sim 100$~billion years, another flyby will be close enough to strip Jupiter from the white dwarf Sun~\citep{zink20}.  This will leave our Sun as a solitary white dwarf.

With no remaining planets orbiting the (white dwarf) Sun, this will mark the end of the Solar System (although, of course, we cannot rule out the possibility that the Sun may capture a new planet or two during these flybys).

\section{Summary and discussion}\label{sec:conclusions}

This chapter has presented our current understanding of the Solar System's formation and evolution.  This story spans 100 billion years in duration, roughly seven times the current age of the Universe.  By necessity, we have used a variety of means to extrapolate what we think must have taken place.  This extrapolation came in the form of scientific models, like the ones presented for the formation of the rocky planets in \S~\ref{sec:rocky}.  

With so many different models, how are we to know which to trust? This is a universal challenge in science.  It is sometimes difficult to remember that no model or idea can ever be proven; it can only be disproved.  The first step in testing models is to evaluate the assumptions that are made. For example, in the case of the terrestrial planet formation models from \S~\ref{sec:rocky}, we can use better computer simulations or astronomical observations to evaluate where and when planetesimals form.  If we learned, for example, that planetesimals do not form in rings, then the starting conditions of the {\em Empty primordial asteroid belt} model would be disproved and that model would be rejected.  Another way to test models is to confront them with some form of ground truth. For instance, the {\em Early Instability} model would be rejected if new cosmochemical data demonstrated conclusively that the giant planet instability took place later than about 20 million years after the start of planet formation. 

This process acts to guide the field of Solar System formation.  Given the many different models that exist, a large portion of the community's effort is dedicated to confronting those models with reality in some form. The search for this reality can involve deep dives into the physics of complex processes such as planetesimal formation or orbital migration.  It can also involve measurements that seek to determine a specific quantity that is central to models, such as the orbits of the giant planets at very early times, or the timing of the Moon-forming impact. This type of approach to choosing research questions is validated by the community, because it builds naturally on previous generations of models.  At the same time, new technologies such as larger telescopes and better instruments drive an exploratory phase that essentially represent new chapters in the Solar System's story. The new observations or measurements must then be put into context using new, more detailed (or, sometimes, completely different) models.  Those models have their own uncertainties and bottlenecks, and the process continues.


In this way, the field of Solar System formation has taken giant strides in recent years.  Yet there are still plenty of important unanswered questions.  For instance, how exactly can we use our understanding of Solar System formation to get a handle on the formation of planets in a more generic, cosmic context? Why is our system structured the way it is, and in what ways does its structure matter? This type of thinking is underway~\citep{raymond22b}, but the vast diversity of exoplanet systems~\citep{winn15} makes it a challenge to have an all-encompassing framework for planet formation. It is also sometimes difficult to remove the anthropic bias from our thinking.  For instance, it can be demonstrated quite simply that the Solar System is an unusual planetary system; if the Sun were explored as an exoplanet host star, Jupiter is the only planet that would be detectable to date.  Yet Jupiter is in a 1\% minority, as only $\sim 10\%$ of Sun-like stars host Jupiter-mass planets, and only $\sim 10\%$ of those have orbits similar to our Jupiter's~\citep{raymond22b}.  Given that we are in a statistically unusual system, how can we ever hope to have a truly unbiased view of our system's cosmic place in the Universe?  It is a challenge to come up with firm answers to these questions.  Nonetheless, asking them acts as a strong motivator to learn more about our place in the Universe, and the history of our Solar System.

\begin{ack}[Acknowledgments]

I am grateful to a few key influences who helped me to become an astrophysicist: my parents, Christophe Roux, my 4th grade teacher at Ecolint, Jim Turner and Roy LaCasce at Bowdoin College, Suzanne Hawley, Vikki Meadows, and Tom Quinn at the University of Washington, Alessandro Morbidelli, and my wife, Marisa.  

I acknowledge funding from a variety of sources during my career, including the NASA Astrobiology Institute (via the University of Washington, University of Colorado, and Virtual Planetary Laboratory lead teams), the NASA Postdoctoral Program (NPP) Felloswhip program, NASA's Origins of Solar Systems program, the Agence Nationale pour la Recherche (ANR), the CNRS' PNP, EPOV, and MITI programs, the PEPR Origins, and the Investments for the Future programme IdEx, Universit{\'e} de Bordeaux/RRI ORIGINS.  
\end{ack}

\seealso{There are many reviews on different phases of Solar System formation and evolution, including \cite{adams10} on the Sun's birth cluster, \cite{williams11} and \cite{armitage11} on the properties and dynamics of protoplanetary disks, \cite{birnstiel16} and \cite{johansen14} on dust growth and drift and the formation of planetesimals, \cite{johansen17} on the growth of planetary embryos and giant planets by pebble and planetesimal accretion, \cite{kley12} on orbital migration, \cite{nesvorny18} on the giant planet instability, \cite{raymond22b} on terrestrial planet formation models, and \cite{canup21} on the Moon-forming impact, \cite{raymondnesvorny22} on the origins and dynamics of the asteroid belt. There are also excellent reviews on planetary formation with a focus on protoplanetary disk- and exoplanet observations~\citep{drazkowska23}, the meteorite isotope dichotomy~\citep{kleine20}, the properties of known exoplanet systems~\citep{winn15}, the origin of Earth's composition~\citep{krijt23}, and \cite{veras16} on the evolution of planetary systems during and after their star's post-main sequence evolution.}


\begin{thebibliography*}{42}
\providecommand{\bibtype}[1]{}
\providecommand{\natexlab}[1]{#1}
{\catcode`\|=0\catcode`\#=12\catcode`\@=11\catcode`\\=12
|immediate|write|@auxout{\expandafter\ifx\csname
  natexlab\endcsname\relax\gdef\natexlab#1{#1}\fi}}
\renewcommand{\url}[1]{{\tt #1}}
\providecommand{\urlprefix}{URL }
\expandafter\ifx\csname urlstyle\endcsname\relax
  \providecommand{\doi}[1]{doi:\discretionary{}{}{}#1}\else
  \providecommand{\doi}{doi:\discretionary{}{}{}\begingroup
  \urlstyle{rm}\Url}\fi
\providecommand{\bibinfo}[2]{#2}
\providecommand{\eprint}[2][]{\url{#2}}

\bibtype{Article}%
\bibitem[{Adams}(2010)]{adams10}
\bibinfo{author}{{Adams} FC} (\bibinfo{year}{2010}), \bibinfo{month}{Sep.}
\bibinfo{title}{{The Birth Environment of the Solar System}}.
\bibinfo{journal}{{\em \araa}} \bibinfo{volume}{48}: \bibinfo{pages}{47--85}.
  \bibinfo{doi}{\doi{10.1146/annurev-astro-081309-130830}}.
\eprint{1001.5444}.

\bibtype{Article}%
\bibitem[{Armitage}(2011)]{armitage11}
\bibinfo{author}{{Armitage} PJ} (\bibinfo{year}{2011}), \bibinfo{month}{Sep.}
\bibinfo{title}{{Dynamics of Protoplanetary Disks}}.
\bibinfo{journal}{{\em \araa}} \bibinfo{volume}{49}: \bibinfo{pages}{195--236}.
  \bibinfo{doi}{\doi{10.1146/annurev-astro-081710-102521}}.
\eprint{1011.1496}.

\bibtype{Article}%
\bibitem[{Batygin} et al.(2019)]{batygin19}
\bibinfo{author}{{Batygin} K}, \bibinfo{author}{{Adams} FC},
  \bibinfo{author}{{Brown} ME} and  \bibinfo{author}{{Becker} JC}
  (\bibinfo{year}{2019}), \bibinfo{month}{May}.
\bibinfo{title}{{The planet nine hypothesis}}.
\bibinfo{journal}{{\em Physics Reports}} \bibinfo{volume}{805}:
  \bibinfo{pages}{1--53}. \bibinfo{doi}{\doi{10.1016/j.physrep.2019.01.009}}.
\eprint{1902.10103}.

\bibtype{Article}%
\bibitem[{Birnstiel} et al.(2016)]{birnstiel16}
\bibinfo{author}{{Birnstiel} T}, \bibinfo{author}{{Fang} M} and
  \bibinfo{author}{{Johansen} A} (\bibinfo{year}{2016}), \bibinfo{month}{Dec.}
\bibinfo{title}{{Dust Evolution and the Formation of Planetesimals}}.
\bibinfo{journal}{{\em \ssr}} \bibinfo{volume}{205}: \bibinfo{pages}{41--75}.
  \bibinfo{doi}{\doi{10.1007/s11214-016-0256-1}}.
\eprint{1604.02952}.

\bibtype{Article}%
\bibitem[{Bottke} and {Norman}(2017)]{bottke17}
\bibinfo{author}{{Bottke} WF} and  \bibinfo{author}{{Norman} MD}
  (\bibinfo{year}{2017}), \bibinfo{month}{Aug.}
\bibinfo{title}{{The Late Heavy Bombardment}}.
\bibinfo{journal}{{\em Annual Review of Earth and Planetary Sciences}}
  \bibinfo{volume}{45} (\bibinfo{number}{1}): \bibinfo{pages}{619--647}.
  \bibinfo{doi}{\doi{10.1146/annurev-earth-063016-020131}}.

\bibtype{Article}%
\bibitem[{Bro{\v{z}}} et al.(2021)]{broz21}
\bibinfo{author}{{Bro{\v{z}}} M}, \bibinfo{author}{{Chrenko} O},
  \bibinfo{author}{{Nesvorn{\'y}} D} and  \bibinfo{author}{{Dauphas} N}
  (\bibinfo{year}{2021}), \bibinfo{month}{Jul.}
\bibinfo{title}{{Early terrestrial planet formation by torque-driven convergent
  migration of planetary embryos}}.
\bibinfo{journal}{{\em Nature Astronomy}} \bibinfo{volume}{5}:
  \bibinfo{pages}{898--902}. \bibinfo{doi}{\doi{10.1038/s41550-021-01383-3}}.
\eprint{2109.11385}.

\bibtype{Article}%
\bibitem[{Canup} et al.(2021)]{canup21}
\bibinfo{author}{{Canup} RM}, \bibinfo{author}{{Righter} K},
  \bibinfo{author}{{Dauphas} N}, \bibinfo{author}{{Pahlevan} K},
  \bibinfo{author}{{{\'C}uk} M}, \bibinfo{author}{{Lock} SJ},
  \bibinfo{author}{{Stewart} ST}, \bibinfo{author}{{Salmon} J},
  \bibinfo{author}{{Rufu} R}, \bibinfo{author}{{Nakajima} M} and
  \bibinfo{author}{{Magna} T} (\bibinfo{year}{2021}), \bibinfo{month}{Mar.}
\bibinfo{title}{{Origin of the Moon}}.
\bibinfo{journal}{{\em arXiv e-prints}} ,
  \bibinfo{eid}{arXiv:2103.02045}\bibinfo{doi}{\doi{10.48550/arXiv.2103.02045}}.
\eprint{2103.02045}.

\bibtype{Article}%
\bibitem[{Clement} et al.(2018)]{clement18}
\bibinfo{author}{{Clement} MS}, \bibinfo{author}{{Kaib} NA},
  \bibinfo{author}{{Raymond} SN} and  \bibinfo{author}{{Walsh} KJ}
  (\bibinfo{year}{2018}), \bibinfo{month}{Sep}.
\bibinfo{title}{{Mars' growth stunted by an early giant planet instability}}.
\bibinfo{journal}{{\em \icarus}} \bibinfo{volume}{311}:
  \bibinfo{pages}{340--356}. \bibinfo{doi}{\doi{10.1016/j.icarus.2018.04.008}}.
\eprint{1804.04233}.

\bibtype{Article}%
\bibitem[{de Sousa} et al.(2020)]{ribeiro20}
\bibinfo{author}{{de Sousa} RR}, \bibinfo{author}{{Morbidelli} A},
  \bibinfo{author}{{Raymond} SN}, \bibinfo{author}{{Izidoro} A},
  \bibinfo{author}{{Gomes} R} and  \bibinfo{author}{{Vieira Neto} E}
  (\bibinfo{year}{2020}), \bibinfo{month}{Mar.}
\bibinfo{title}{{Dynamical evidence for an early giant planet instability}}.
\bibinfo{journal}{{\em \icarus}} \bibinfo{volume}{339}, \bibinfo{eid}{113605}.
  \bibinfo{doi}{\doi{10.1016/j.icarus.2019.113605}}.

\bibtype{Article}%
\bibitem[{DeMeo} and {Carry}(2013)]{demeo13}
\bibinfo{author}{{DeMeo} FE} and  \bibinfo{author}{{Carry} B}
  (\bibinfo{year}{2013}), \bibinfo{month}{Sep.}
\bibinfo{title}{{The taxonomic distribution of asteroids from multi-filter
  all-sky photometric surveys}}.
\bibinfo{journal}{{\em Icarus}} \bibinfo{volume}{226}:
  \bibinfo{pages}{723--741}. \bibinfo{doi}{\doi{10.1016/j.icarus.2013.06.027}}.
\eprint{1307.2424}.

\bibtype{Inproceedings}%
\bibitem[{Dra{\.z}kowska} et al.(2023)]{drazkowska23}
\bibinfo{author}{{Dra{\.z}kowska} J}, \bibinfo{author}{{Bitsch} B},
  \bibinfo{author}{{Lambrechts} M}, \bibinfo{author}{{Mulders} GD},
  \bibinfo{author}{{Harsono} D}, \bibinfo{author}{{Vazan} A},
  \bibinfo{author}{{Liu} B}, \bibinfo{author}{{Ormel} CW},
  \bibinfo{author}{{Kretke} K} and  \bibinfo{author}{{Morbidelli} A}
  (\bibinfo{year}{2023}), \bibinfo{month}{Jul.}, \bibinfo{title}{{Planet
  Formation Theory in the Era of ALMA and Kepler: from Pebbles to Exoplanets}},
  \bibinfo{editor}{{Inutsuka} S}, \bibinfo{editor}{{Aikawa} Y},
  \bibinfo{editor}{{Muto} T}, \bibinfo{editor}{{Tomida} K} and
  \bibinfo{editor}{{Tamura} M}, (Eds.), \bibinfo{booktitle}{Protostars and
  Planets VII}, \bibinfo{series}{Astronomical Society of the Pacific Conference
  Series}, \bibinfo{volume}{534}, pp. \bibinfo{pages}{717},
  \eprint{2203.09759}.

\bibtype{Article}%
\bibitem[{Fernandez} and {Ip}(1984)]{fernandez84}
\bibinfo{author}{{Fernandez} JA} and  \bibinfo{author}{{Ip} W}
  (\bibinfo{year}{1984}), \bibinfo{month}{Apr.}
\bibinfo{title}{{Some dynamical aspects of the accretion of Uranus and Neptune
  - The exchange of orbital angular momentum with planetesimals}}.
\bibinfo{journal}{{\em Icarus}} \bibinfo{volume}{58}:
  \bibinfo{pages}{109--120}. \bibinfo{doi}{\doi{10.1016/0019-1035(84)90101-5}}.

\bibtype{Article}%
\bibitem[{Fraser} et al.(2022)]{fraser22}
\bibinfo{author}{{Fraser} WC}, \bibinfo{author}{{Dones} L},
  \bibinfo{author}{{Volk} K}, \bibinfo{author}{{Womack} M} and
  \bibinfo{author}{{Nesvorn{\'y}} D} (\bibinfo{year}{2022}),
  \bibinfo{month}{Oct.}
\bibinfo{title}{{The Transition from the Kuiper Belt to the Jupiter-Family
  (Comets)}}.
\bibinfo{journal}{{\em arXiv e-prints}} ,
  \bibinfo{eid}{arXiv:2210.16354}\bibinfo{doi}{\doi{10.48550/arXiv.2210.16354}}.
\eprint{2210.16354}.

\bibtype{Article}%
\bibitem[{Gomes} et al.(2005)]{gomes05}
\bibinfo{author}{{Gomes} R}, \bibinfo{author}{{Levison} HF},
  \bibinfo{author}{{Tsiganis} K} and  \bibinfo{author}{{Morbidelli} A}
  (\bibinfo{year}{2005}), \bibinfo{month}{May}.
\bibinfo{title}{{Origin of the cataclysmic Late Heavy Bombardment period of the
  terrestrial planets}}.
\bibinfo{journal}{{\em \nat}} \bibinfo{volume}{435}: \bibinfo{pages}{466--469}.
  \bibinfo{doi}{\doi{10.1038/nature03676}}.

\bibtype{Article}%
\bibitem[{Hays} et al.(1976)]{hays76}
\bibinfo{author}{{Hays} JD}, \bibinfo{author}{{Imbrie} J} and
  \bibinfo{author}{{Shackleton} NJ} (\bibinfo{year}{1976}),
  \bibinfo{month}{Dec.}
\bibinfo{title}{{Variations in the Earth's Orbit: Pacemaker of the Ice Ages}}.
\bibinfo{journal}{{\em Science}} \bibinfo{volume}{194}
  (\bibinfo{number}{4270}): \bibinfo{pages}{1121--1132}.
  \bibinfo{doi}{\doi{10.1126/science.194.4270.1121}}.

\bibtype{Article}%
\bibitem[{Izidoro} et al.(2022)]{izidoro22}
\bibinfo{author}{{Izidoro} A}, \bibinfo{author}{{Dasgupta} R},
  \bibinfo{author}{{Raymond} SN}, \bibinfo{author}{{Deienno} R},
  \bibinfo{author}{{Bitsch} B} and  \bibinfo{author}{{Isella} A}
  (\bibinfo{year}{2022}), \bibinfo{month}{Mar.}
\bibinfo{title}{{Planetesimal rings as the cause of the Solar System's
  planetary architecture}}.
\bibinfo{journal}{{\em Nature Astronomy}} \bibinfo{volume}{6}:
  \bibinfo{pages}{357--366}. \bibinfo{doi}{\doi{10.1038/s41550-021-01557-z}}.
\eprint{2112.15558}.

\bibtype{Article}%
\bibitem[{Johansen} and {Lambrechts}(2017)]{johansen17}
\bibinfo{author}{{Johansen} A} and  \bibinfo{author}{{Lambrechts} M}
  (\bibinfo{year}{2017}), \bibinfo{month}{Aug.}
\bibinfo{title}{{Forming Planets via Pebble Accretion}}.
\bibinfo{journal}{{\em Annual Review of Earth and Planetary Sciences}}
  \bibinfo{volume}{45}: \bibinfo{pages}{359--387}.
  \bibinfo{doi}{\doi{10.1146/annurev-earth-063016-020226}}.

\bibtype{Article}%
\bibitem[{Johansen} et al.(2014)]{johansen14}
\bibinfo{author}{{Johansen} A}, \bibinfo{author}{{Blum} J},
  \bibinfo{author}{{Tanaka} H}, \bibinfo{author}{{Ormel} C},
  \bibinfo{author}{{Bizzarro} M} and  \bibinfo{author}{{Rickman} H}
  (\bibinfo{year}{2014}).
\bibinfo{title}{{The Multifaceted Planetesimal Formation Process}}.
\bibinfo{journal}{{\em Protostars and Planets VI}} :
  \bibinfo{pages}{547--570}\eprint{1402.1344}.

\bibtype{Article}%
\bibitem[{Johansen} et al.(2021)]{johansen21}
\bibinfo{author}{{Johansen} A}, \bibinfo{author}{{Ronnet} T},
  \bibinfo{author}{{Bizzarro} M}, \bibinfo{author}{{Schiller} M},
  \bibinfo{author}{{Lambrechts} M}, \bibinfo{author}{{Nordlund} {\r{A}}} and
  \bibinfo{author}{{Lammer} H} (\bibinfo{year}{2021}), \bibinfo{month}{Feb.}
\bibinfo{title}{{A pebble accretion model for the formation of the terrestrial
  planets in the Solar System}}.
\bibinfo{journal}{{\em Science Advances}} \bibinfo{volume}{7}
  (\bibinfo{number}{8}): \bibinfo{pages}{eabc0444}.
  \bibinfo{doi}{\doi{10.1126/sciadv.abc0444}}.
\eprint{2102.08611}.

\bibtype{Article}%
\bibitem[{Kaib} and {Volk}(2022)]{kaib22}
\bibinfo{author}{{Kaib} NA} and  \bibinfo{author}{{Volk} K}
  (\bibinfo{year}{2022}), \bibinfo{month}{May}.
\bibinfo{title}{{Dynamical Population of Comet Reservoirs}}.
\bibinfo{journal}{{\em arXiv e-prints}} ,
  \bibinfo{eid}{arXiv:2206.00010}\bibinfo{doi}{\doi{10.48550/arXiv.2206.00010}}.
\eprint{2206.00010}.

\bibtype{Article}%
\bibitem[{Kleine} and {Walker}(2017)]{kleine17}
\bibinfo{author}{{Kleine} T} and  \bibinfo{author}{{Walker} RJ}
  (\bibinfo{year}{2017}), \bibinfo{month}{Aug.}
\bibinfo{title}{{Tungsten Isotopes in Planets}}.
\bibinfo{journal}{{\em Annual Review of Earth and Planetary Sciences}}
  \bibinfo{volume}{45} (\bibinfo{number}{1}): \bibinfo{pages}{389--417}.
  \bibinfo{doi}{\doi{10.1146/annurev-earth-063016-020037}}.

\bibtype{Article}%
\bibitem[{Kleine} et al.(2020)]{kleine20}
\bibinfo{author}{{Kleine} T}, \bibinfo{author}{{Budde} G},
  \bibinfo{author}{{Burkhardt} C}, \bibinfo{author}{{Kruijer} TS},
  \bibinfo{author}{{Worsham} EA}, \bibinfo{author}{{Morbidelli} A} and
  \bibinfo{author}{{Nimmo} F} (\bibinfo{year}{2020}), \bibinfo{month}{May}.
\bibinfo{title}{{The Non-carbonaceous-Carbonaceous Meteorite Dichotomy}}.
\bibinfo{journal}{{\em \ssr}} \bibinfo{volume}{216} (\bibinfo{number}{4}),
  \bibinfo{eid}{55}. \bibinfo{doi}{\doi{10.1007/s11214-020-00675-w}}.

\bibtype{Article}%
\bibitem[{Kley} and {Nelson}(2012)]{kley12}
\bibinfo{author}{{Kley} W} and  \bibinfo{author}{{Nelson} RP}
  (\bibinfo{year}{2012}), \bibinfo{month}{Sep.}
\bibinfo{title}{{Planet-Disk Interaction and Orbital Evolution}}.
\bibinfo{journal}{{\em \araa}} \bibinfo{volume}{50}: \bibinfo{pages}{211--249}.
  \bibinfo{doi}{\doi{10.1146/annurev-astro-081811-125523}}.
\eprint{1203.1184}.

\bibtype{Inproceedings}%
\bibitem[{Krijt} et al.(2023)]{krijt23}
\bibinfo{author}{{Krijt} S}, \bibinfo{author}{{Kama} M},
  \bibinfo{author}{{McClure} M}, \bibinfo{author}{{Teske} J},
  \bibinfo{author}{{Bergin} EA}, \bibinfo{author}{{Shorttle} O},
  \bibinfo{author}{{Walsh} KJ} and  \bibinfo{author}{{Raymond} SN}
  (\bibinfo{year}{2023}), \bibinfo{month}{Jul.}, \bibinfo{title}{{Chemical
  Habitability: Supply and Retention of Life's Essential Elements During Planet
  Formation}}, \bibinfo{editor}{{Inutsuka} S}, \bibinfo{editor}{{Aikawa} Y},
  \bibinfo{editor}{{Muto} T}, \bibinfo{editor}{{Tomida} K} and
  \bibinfo{editor}{{Tamura} M}, (Eds.), \bibinfo{booktitle}{Protostars and
  Planets VII}, \bibinfo{series}{Astronomical Society of the Pacific Conference
  Series}, \bibinfo{volume}{534}, pp. \bibinfo{pages}{1031}.

\bibtype{Article}%
\bibitem[{Laskar}(1994)]{laskar94}
\bibinfo{author}{{Laskar} J} (\bibinfo{year}{1994}), \bibinfo{month}{Jul.}
\bibinfo{title}{{Large-scale chaos in the solar system.}}
\bibinfo{journal}{{\em \aap}} \bibinfo{volume}{287}: \bibinfo{pages}{L9--L12}.

\bibtype{Article}%
\bibitem[{Liu} et al.(2022)]{liu22}
\bibinfo{author}{{Liu} B}, \bibinfo{author}{{Raymond} SN} and
  \bibinfo{author}{{Jacobson} SA} (\bibinfo{year}{2022}), \bibinfo{month}{Apr.}
\bibinfo{title}{{Early Solar System instability triggered by dispersal of the
  gaseous disk}}.
\bibinfo{journal}{{\em \nat}} \bibinfo{volume}{604} (\bibinfo{number}{7907}):
  \bibinfo{pages}{643--646}. \bibinfo{doi}{\doi{10.1038/s41586-022-04535-1}}.
\eprint{2205.02026}.

\bibtype{Article}%
\bibitem[{Lykawka} and {Ito}(2023)]{lykawka23}
\bibinfo{author}{{Lykawka} PS} and  \bibinfo{author}{{Ito} T}
  (\bibinfo{year}{2023}), \bibinfo{month}{Mar.}
\bibinfo{title}{{Terrestrial planet and asteroid belt formation by
  Jupiter-Saturn chaotic excitation}}.
\bibinfo{journal}{{\em Scientific Reports}} \bibinfo{volume}{13},
  \bibinfo{eid}{4708}. \bibinfo{doi}{\doi{10.1038/s41598-023-30382-9}}.
\eprint{2208.13647}.

\bibtype{Article}%
\bibitem[{Malhotra}(1993)]{malhotra93}
\bibinfo{author}{{Malhotra} R} (\bibinfo{year}{1993}), \bibinfo{month}{Oct.}
\bibinfo{title}{{The origin of Pluto's peculiar orbit}}.
\bibinfo{journal}{{\em \nat}} \bibinfo{volume}{365}: \bibinfo{pages}{819--821}.
  \bibinfo{doi}{\doi{10.1038/365819a0}}.

\bibtype{Article}%
\bibitem[{Nesvorn{\'y}} et al.(2018)]{nesvorny18}
\bibinfo{author}{{Nesvorn{\'y}} D}, \bibinfo{author}{{Vokrouhlick{\'y}} D},
  \bibinfo{author}{{Bottke} WF} and  \bibinfo{author}{{Levison} HF}
  (\bibinfo{year}{2018}), \bibinfo{month}{Sep.}
\bibinfo{title}{{Evidence for very early migration of the Solar System planets
  from the Patroclus-Menoetius binary Jupiter Trojan}}.
\bibinfo{journal}{{\em Nature Astronomy}} \bibinfo{volume}{2}:
  \bibinfo{pages}{878--882}. \bibinfo{doi}{\doi{10.1038/s41550-018-0564-3}}.
\eprint{1809.04007}.

\bibtype{Article}%
\bibitem[{Quinn} et al.(1991)]{quinn91}
\bibinfo{author}{{Quinn} TR}, \bibinfo{author}{{Tremaine} S} and
  \bibinfo{author}{{Duncan} M} (\bibinfo{year}{1991}), \bibinfo{month}{Jun.}
\bibinfo{title}{{A three million year integration of the earth's orbit}}.
\bibinfo{journal}{{\em \aj}} \bibinfo{volume}{101}:
  \bibinfo{pages}{2287--2305}. \bibinfo{doi}{\doi{10.1086/115850}}.

\bibtype{Inproceedings}%
\bibitem[{Raymond} and {Morbidelli}(2022)]{raymond22b}
\bibinfo{author}{{Raymond} SN} and  \bibinfo{author}{{Morbidelli} A}
  (\bibinfo{year}{2022}), \bibinfo{month}{Jan.}, \bibinfo{title}{{Planet
  Formation: Key Mechanisms and Global Models}}, \bibinfo{editor}{{Biazzo} K},
  \bibinfo{editor}{{Bozza} V}, \bibinfo{editor}{{Mancini} L} and
  \bibinfo{editor}{{Sozzetti} A}, (Eds.), \bibinfo{booktitle}{Demographics of
  Exoplanetary Systems, Lecture Notes of the 3rd Advanced School on
  Exoplanetary Science}, \bibinfo{series}{Astrophysics and Space Science
  Library}, \bibinfo{volume}{466},  \bibinfo{pages}{3--82},
  \eprint{2002.05756}.

\bibtype{incollection}%
\bibitem[{Raymond} and {Nesvorn{\'y}}(2022)]{raymondnesvorny22}
\bibinfo{author}{{Raymond} SN} and  \bibinfo{author}{{Nesvorn{\'y}} D}
  (\bibinfo{year}{2022}), \bibinfo{title}{{Origin and Dynamical Evolution of
  the Asteroid Belt}}, \bibinfo{booktitle}{Vesta and Ceres. Insights from the
  Dawn Mission for the Origin of the Solar System}, pp. \bibinfo{pages}{227}.

\bibtype{Article}%
\bibitem[{Steller} et al.(2022)]{steller22}
\bibinfo{author}{{Steller} T}, \bibinfo{author}{{Burkhardt} C},
  \bibinfo{author}{{Yang} C} and  \bibinfo{author}{{Kleine} T}
  (\bibinfo{year}{2022}), \bibinfo{month}{Nov.}
\bibinfo{title}{{Nucleosynthetic zinc isotope anomalies reveal a dual origin of
  terrestrial volatiles}}.
\bibinfo{journal}{{\em \icarus}} \bibinfo{volume}{386}, \bibinfo{eid}{115171}.
  \bibinfo{doi}{\doi{10.1016/j.icarus.2022.115171}}.

\bibtype{Article}%
\bibitem[{Veras}(2016)]{veras16}
\bibinfo{author}{{Veras} D} (\bibinfo{year}{2016}), \bibinfo{month}{Feb.}
\bibinfo{title}{{Post-main-sequence planetary system evolution}}.
\bibinfo{journal}{{\em Royal Society Open Science}} \bibinfo{volume}{3},
  \bibinfo{eid}{150571}. \bibinfo{doi}{\doi{10.1098/rsos.150571}}.
\eprint{1601.05419}.

\bibtype{incollection}%
\bibitem[{Vokrouhlick{\'y}} et al.(2015)]{vok15}
\bibinfo{author}{{Vokrouhlick{\'y}} D}, \bibinfo{author}{{Bottke} WF},
  \bibinfo{author}{{Chesley} SR}, \bibinfo{author}{{Scheeres} DJ} and
  \bibinfo{author}{{Statler} TS} (\bibinfo{year}{2015}), \bibinfo{title}{{The
  Yarkovsky and YORP Effects}}, \bibinfo{booktitle}{Asteroids IV},
  \bibinfo{pages}{509--531}.

\bibtype{Article}%
\bibitem[{Walker}(2009)]{walker09}
\bibinfo{author}{{Walker} RJ} (\bibinfo{year}{2009}), \bibinfo{month}{Jun.}
\bibinfo{title}{{Highly siderophile elements in the Earth, Moon and Mars:
  Update and implications for planetary accretion and differentiation}}.
\bibinfo{journal}{{\em Chemie der Erde / Geochemistry}} \bibinfo{volume}{69}:
  \bibinfo{pages}{101--125}. \bibinfo{doi}{\doi{10.1016/j.chemer.2008.10.001}}.

\bibtype{Article}%
\bibitem[{Walsh} et al.(2011)]{walsh11}
\bibinfo{author}{{Walsh} KJ}, \bibinfo{author}{{Morbidelli} A},
  \bibinfo{author}{{Raymond} SN}, \bibinfo{author}{{O'Brien} DP} and
  \bibinfo{author}{{Mandell} AM} (\bibinfo{year}{2011}), \bibinfo{month}{Jul.}
\bibinfo{title}{{A low mass for Mars from Jupiter's early gas-driven
  migration}}.
\bibinfo{journal}{{\em \nat}} \bibinfo{volume}{475}: \bibinfo{pages}{206--209}.
  \bibinfo{doi}{\doi{10.1038/nature10201}}.

\bibtype{Inproceedings}%
\bibitem[{Wetherill}(1991)]{wetherill91}
\bibinfo{author}{{Wetherill} GW} (\bibinfo{year}{1991}), \bibinfo{month}{Mar.},
  \bibinfo{title}{{Why Isn't Mars as Big as Earth?}}, \bibinfo{booktitle}{Lunar
  and Planetary Institute Science Conference Abstracts}, \bibinfo{series}{Lunar
  and Planetary Inst. Technical Report}, \bibinfo{volume}{22}, pp.
  \bibinfo{pages}{1495}.

\bibtype{Article}%
\bibitem[{Williams} and {Cieza}(2011)]{williams11}
\bibinfo{author}{{Williams} JP} and  \bibinfo{author}{{Cieza} LA}
  (\bibinfo{year}{2011}), \bibinfo{month}{Sep.}
\bibinfo{title}{{Protoplanetary Disks and Their Evolution}}.
\bibinfo{journal}{{\em \araa}} \bibinfo{volume}{49}: \bibinfo{pages}{67--117}.
  \bibinfo{doi}{\doi{10.1146/annurev-astro-081710-102548}}.
\eprint{1103.0556}.

\bibtype{Article}%
\bibitem[{Winn} and {Fabrycky}(2015)]{winn15}
\bibinfo{author}{{Winn} JN} and  \bibinfo{author}{{Fabrycky} DC}
  (\bibinfo{year}{2015}), \bibinfo{month}{Aug.}
\bibinfo{title}{{The Occurrence and Architecture of Exoplanetary Systems}}.
\bibinfo{journal}{{\em \araa}} \bibinfo{volume}{53}: \bibinfo{pages}{409--447}.
  \bibinfo{doi}{\doi{10.1146/annurev-astro-082214-122246}}.
\eprint{1410.4199}.

\bibtype{Article}%
\bibitem[{Zellner}(2017)]{zellner17}
\bibinfo{author}{{Zellner} NEB} (\bibinfo{year}{2017}), \bibinfo{month}{Sep.}
\bibinfo{title}{{Cataclysm No More: New Views on the Timing and Delivery of
  Lunar Impactors}}.
\bibinfo{journal}{{\em Origins of Life and Evolution of the Biosphere}}
  \bibinfo{volume}{47}: \bibinfo{pages}{261--280}.
  \bibinfo{doi}{\doi{10.1007/s11084-017-9536-3}}.
\eprint{1704.06694}.

\bibtype{Article}%
\bibitem[{Zink} et al.(2020)]{zink20}
\bibinfo{author}{{Zink} JK}, \bibinfo{author}{{Batygin} K} and
  \bibinfo{author}{{Adams} FC} (\bibinfo{year}{2020}), \bibinfo{month}{Nov.}
\bibinfo{title}{{The Great Inequality and the Dynamical Disintegration of the
  Outer Solar System}}.
\bibinfo{journal}{{\em \aj}} \bibinfo{volume}{160} (\bibinfo{number}{5}),
  \bibinfo{eid}{232}. \bibinfo{doi}{\doi{10.3847/1538-3881/abb8de}}.
\eprint{2009.07296}.

\end{thebibliography*}

\end{document}